\pdfoutput=1

\documentclass[twocolumn, times]{aastex61}

\hypersetup{linkcolor=red,citecolor=blue,filecolor=cyan,urlcolor=magenta}

\received{}
\revised{}
\accepted{}

\submitjournal{ApJ}

\shorttitle{Gas kinematics in a protocluster at z=2.5}
\shortauthors{Minju M. Lee et al.}
\defcitealias{minju2017a}{Paper I}

\begin{document}

\title{A Radio-to-millimeter Census of Star-forming Galaxies in Protocluster 4C~23.56 at z = 2.5 : Global and local gas kinematics}

\correspondingauthor{Minju M. Lee}
\email{minju@mpe.mpg.de}

\author[0000-0002-2419-3068]{Minju M. Lee}
\affiliation{Max-Planck-Institut f\"{u}r Extraterrestrische Physik (MPE), Giessenbachstr., D-85748 Garching, Germany}
\affiliation{Division of Particle and Astrophysical Science, Graduate School of Science, Nagoya University, Furo-cho, Chikusa-ku, Nagoya 464-8602, Japan}
\affiliation{National Observatory of Japan, 2-21-1 Osawa, Mitaka, Tokyo 181-0015, Japan}

\author{Ichi Tanaka}
\affiliation{Subaru Telescope, National Astronomical Observatory of Japan, 650 North Aohoku Place, Hilo, HI 96720, USA}

\author{Ryohei Kawabe}
\affiliation{Department of Astronomy, The University of Tokyo, 7-3-1 Hongo, Bunkyo-ku, Tokyo 133-0033, Japan}
\affiliation{National Observatory of Japan, 2-21-1 Osawa, Mitaka, Tokyo 181-0015, Japan}
\affiliation{SOKENDAI (The Graduate University for Advanced Studies), 2-21-1 Osawa, Mitaka, Tokyo 181-0015, Japan}

\author{Itziar Aretxaga}
\affiliation{Instituto Nacional de Astrofisica, Optica y Electronica (INAOE), Aptdo. Postal 51 y 216, 72000 Puebla, Mexico}

\author{Bunyo Hatsukade}
\affiliation{Institute of Astronomy, The University of Tokyo, 2-21-1 Osawa, Mitaka, Tokyo, 181-0015, Japan}

\author{Takuma Izumi}
\affiliation{National Observatory of Japan, 2-21-1 Osawa, Mitaka, Tokyo 181-0015, Japan}

\author{Masaru Kajisawa}
\affiliation{Graduate School of Science and Engineering, Ehime University, Bunkyo-cho, Matsuyama 790-8577, Japan}
\affiliation{Research Center for Space and Cosmic Evolution, Ehime University, Bunkyo-cho, Matsuyama 790-8577, Japan}

\author{Tadayuki Kodama}
\affiliation{Astronomical Institute, Tohoku University, Aoba-ku, Sendai 980-8578, Japan}

\author{Kotaro Kohno}
\affiliation{Institute of Astronomy, The University of Tokyo, 2-21-1 Osawa, Mitaka, Tokyo, 181-0015, Japan}
\affiliation{Research Center for the Early Universe, The University of Tokyo, 7-3-1 Hongo, Bunkyo, Tokyo 113-0033, Japan}

\author{Kouichiro Nakanishi}
\affiliation{National Observatory of Japan, 2-21-1 Osawa, Mitaka, Tokyo 181-0015, Japan}
\affiliation{SOKENDAI (The Graduate University for Advanced Studies), 2-21-1 Osawa, Mitaka, Tokyo 181-0015, Japan}

\author{Toshiki Saito}
\affiliation{Department of Astronomy, The University of Tokyo, 7-3-1 Hongo, Bunkyo-ku, Tokyo 133-0033, Japan}
\affiliation{National Observatory of Japan, 2-21-1 Osawa, Mitaka, Tokyo 181-0015, Japan}

\author{Ken-ichi Tadaki}
\affiliation{National Observatory of Japan, 2-21-1 Osawa, Mitaka, Tokyo 181-0015, Japan}

\author{Yoichi Tamura}
\affiliation{Department of Physics, Nagoya University, Furo-cho, Chikusa-ku, Nagoya 464-8601, Japan}

\author{Hideki Umehata}
\affiliation{Institute of Astronomy, The University of Tokyo, 2-21-1 Osawa, Mitaka, Tokyo, 181-0015, Japan}
\affiliation{The Open University of Japan, 2-11 Wakaba, Mihama-ku, Chiba 261-8586, Japan}

\author{Milagros Zeballos}
\affiliation{Universidad de las Am\'{e}ricas Puebla, Ex Hacienda Sta Catarina M\'{a}rtir S/N, San Andr\'{e}s Cholula, Puebla CP 72810, Mexico}



\begin{abstract}
We present a study of the gas kinematics of star-forming galaxies associated with protocluster 4C~23.56 at $z=2.49$ using $0''.4$ resolution CO (4--3) data taken with ALMA. 
Eleven H$\alpha$ emitters (HAEs) are detected in CO~(4--3), including six HAEs that were previously detected in CO~(3--2) at a coarser angular resolution. 
The detections in both CO lines are broadly consistent in the line widths and the redshifts, confirming both detections. With  an increase in the number of spectroscopic redshifts, we confirm that the protocluster is composed of two merging groups with a total halo mass of $\log{(M_{\rm cl}/M_{\odot})} =13.4-13.6$, suggesting that the protocluster would evolve into a Virgo-like cluster ($>10^{14} M_{\odot}$). 
We compare the CO line widths and the CO luminosities with galaxies in other (proto)clusters ($n_{\rm gal}=91$) and general fields ($n_{\rm gal}=80$) from other studies.
The 4C23.56 protocluster galaxies have CO line widths and luminosities comparable to other protocluster galaxies on average. On the other hand, the CO line widths are on average broader by $\approx50\%$ compared to field galaxies, while the median CO luminosities are similar.
The broader line widths can be attributed to both effects of unresolved gas-rich mergers and/or compact gas distribution, which is supported by our limited but decent angular resolution observations and the size estimate of three galaxies. Based on these results, we argue that gas-rich mergers may play a role in the retention of the specific angular momentum to a value similar to that of field populations during cluster assembly, though we need to verify this with a larger number of samples.
\end{abstract}



\section{Introduction}\label{sec:intro}
Understanding the formation mechanism of early-type galaxies (ETGs) is one of the long-standing problems in astronomy.
In general, ETGs in the local universe are characterized by old stellar populations, red colors with a small amount of cold gas and dust, and lack of disks (or bulge-dominated) compared to late-type galaxies (LTGs).
The relative fraction of ETGs to LTGs increases toward higher galaxy number densities, which was recognized four decades ago (\citealt{Dressler1980}). Clusters are, therefore, an ideal laboratory to understand ETG formation.

The color and luminosity function evolution in cluster galaxies is consistent with a model in which cluster galaxies formed in short, intensive bursts of star formation at high redshift ($z\gtrsim2$) and evolved passively thereafter (e.g.  \citealt{Bower1992, Stanford1998, Kodama1997, Blakeslee2006, vanDokkum2007, Eisenhardt2008, Kurk2009, Papovich2010}). 
However, it is still elusive how galaxies cease the star-forming activity and change their appearance and kinematical properties at the same time, if necessary, and how significantly the surrounding environment plays a role in these changes at high redshift in particular. The importance of understanding these was magnified in accordance with the kinematical classification of ETGs into fast rotators and slow rotators (\citealt{Emsellem2007, Cappellari2007}, see also \citealt{Cappellari2016} for a recent review) and with the preferential existence of slow rotators in the densest regions in the local universe (e.g., \citealt{Cappellari2011b, Cappellari2013c, Houghton2013, DEugenio2013, Jimmy2013, Fogarty2014, Scott2014, Veale2017b}; but see also \citealt{Brough2017, Greene2017b} who reported no such trend).

In light of this, look-back studies targeting general fields have provided an intriguing connection between massive star-forming main-sequence (e.g., \citealt{Daddi2007, Noeske2007}) and massive quenched galaxies at $z>1$, in terms of their kinematic properties.
The high-$z$ star-forming galaxies exhibit (1) disk-like kinematics even though the star formation rate (SFR) is an order of magnitude higher than the local counterparts and are (2) less rotation dominated than local star-forming galaxies at similar mass, i.e., the ratio of rotation velocity ($V_{\rm rot}$) and intrinsic velocity dispersion ($\sigma_{0}$) decreases toward higher redshift (e.g. \citealt{ForsterSchreiber2006, ForsterSchreiber2009, FS2018a, Genzel2006, Genzel2008, Shapiro2008, Cresci2009, Law2009, Wright2009, Epinat2009, Epinat2012, Daddi2010, Gnerucci2011a, Gnerucci2011b, Vergani2012, Swinbank2012, NewmanS2013, Sobral2013b, Buitrago2014, Stott2016, Leethochawalit2016, Swinbank2017, Turner2017a, Harrison2017, Mason2017, Johnson2018, Girard2018}).
The latter trend of $V_{\rm rot}/\sigma_{0}$ is consistent with an empirical model in which the time evolution is explained by the combination of increasing gas content toward higher redshifts (e.g. \citealt{Magdis2012, Saintonge2013, Tacconi2013, Tacconi2018, Sargent2014, Scoville2014, Scoville2016, Scoville2017b,  Genzel2015, Schinnerer2016, Darvish2018}) and an assumption of a marginally stable (Toomre $Q$ parameter of $Q\sim1$) disk (e.g., \citealt{Wisnioski2015}).
Meanwhile, quenched galaxies at high redshift have (1) compact sizes (e.g., \citealt{Daddi2005, Trujillo2006, Zirm2007, vanDokkum2008, vanDokkum2015, vanderWel2014}) and are explained by (2) disk-like morphologies and surface brightness profiles (\citealt{McGrath2008, Stockton2008, vanDokkum2008, vanderWel2011, Wuyts2011a}). (3) Quiescent galaxies at high redshift are kinematically more rotation dominated, i.e., have higher $V_{\rm rot}/\sigma$ compared to local quiescent galaxies (\citealt{Belli2017a, Toft2017, Newman2018b}).

All together, these studies have provided two insights into the formation of ETGs. 
First, many quiescent galaxies may have formed from disks of star-forming galaxies, which are assembled from smoothly accreting materials, not via major mergers. 
Second, cold gas serves a crucial role for characterizing the kinematical properties of star-forming galaxies in the early universe, which is closely connected to the cessation of star-forming activity and the kinematical transformation to form progenitors of local ETGs.

Do galaxies in denser environments form and evolve similarly as field galaxies explained above? In other words, are they born as ``disk" galaxies and become quiescent through the same physical processes happening in general fields? Or does the environment play a significant role even at high redshifts? These are the questions we would like to answer eventually.

In the hierarchical structure formation, galaxies form earlier in denser dark matter concentrations because the collapse of baryonic matters in dense regions precedes that in less dense regions (\citealt{Springel2005c, DeLucia2006}). 
If the majority of the ETGs form in the same way as described in general field surveys, one would expect no systematic difference in the gas content and gas kinematics in star-forming galaxies in denser regions.
Otherwise, we may find some different properties of galaxies associated with high-redshift (proto)clusters.

Verifying the formation scenario of ETGs in different environments requires spectroscopic redshift confirmation first and then deeper integration for a detailed investigation. This is why for dense environments, rare but bright galaxies (i.e., actively star-forming well above the main sequence) have been preferentially targeted. But, during the last two years, the number of typical star-forming galaxies has greatly increased with CO detection as a probe of gas content (\citealt{minju2017a, Hayashi2017,  Hayashi2018, Noble2017, Stach2017, Castignani2018, Coogan2018,Wang2018}).

In terms of gas kinematics, however, the statistics is still limited because it requires higher angular resolution observations with sufficient integration.
\cite{Dannerbauer2017} has confirmed that a typical main-sequence galaxy associated with a protocluster $z=2.15$ exhibits an extended rotating disk in CO~(1--0).
Recently, \cite{Noble2019} also showed that the majority (i.e., six of eight) of $z\sim1.6$ cluster galaxies exhibit rotating gas disks from CO~(2--1) line observations. They additionally discussed a potential role for the environment, i.e., gas stripping, based on the size of the CO emission.

From the structural information of the stellar component, both star-forming and quiescent galaxies in the core of high-$z$ clusters are fit with a low S\'{e}rsic index of $n<2$, i.e., a disk-like morphology (e.g., \citealt{Strazzullo2013, DePropris2015}). These photometric data-based studies also hint that there is a transition from disk-like star-forming galaxies to spheroidal passive galaxies in cluster regions. 

Some studies postulated that mergers are a mechanism for quenching and kinematical transformation based on the enhanced merger rates from rest-frame optical images (e.g., \citealt{Hine2016, Coogan2018}; but see, e.g., \citealt{Delahaye2017} for no enhancement). However, more observations with kinematic information are needed to pin down the formation mechanism of ETGs in dense environments and to compare with field results and recent hydrodynamical simulations.

In our first paper, we probed the total gas content (\citealt{minju2017a}, hereafter Paper I) for typical star-forming galaxies selected by H$\alpha$ emission, i.e., H$\alpha$ emitters (HAEs) that are associated with the protocluster 4C~23.56 at $z=2.5$.
We note that \citetalias{minju2017a} is one of the first papers that highlight the gas content of typical star-forming galaxies on the main sequence associated with protoclusters at redshift greater than two.
We used CO~(3--2) line emissions and detected the redshifted CO~(3--2) emissions from 7 out of the 22 HAEs that were targeted.
We found no significant difference in the gas fraction for the average value, i.e., $f_{\rm gas, 4C~23.56} \simeq 0.5$ compared to general fields at similar redshift, but instead found a trend of higher star-forming efficiency (SFE) for more massive galaxies that are located in denser regions. This latter trend has been confirmed by collecting a larger number of protocluster galaxies selected by HAEs (\citealt{Tadaki2019a}). Thus, an additional effect seems to play a role for star-forming galaxies in high-$z$ protoclusters.

In this paper, we aim to provide quantitative measurements of the kinematic properties of typical star-forming galaxies with higher angular resolution imaging by focusing on the same HAEs in the well-characterized protocluster 4C~23.56 at $z=2.5$. We also present the kinematic structure of the protocluster itself with increased spectroscopic confirmation from CO that would additionally help our interpretation of the observed kinematical properties in the context of cosmological mass assembly.

This paper is organized as follows. 
In Section~\ref{sec:obs}, a summary of the CO~(4--3) line observations and data analysis procedures is presented.
Section~\ref{sec:results} presents the detection of CO~(4--3) lines for individual galaxies and a comparison with previous CO~(3--2) detections.
In Section~\ref{sec:halos}, we estimate the halo mass of the protocluster with increased redshift information.
In Section~\ref{sec:kinematics}, we discuss the nature of two gas-rich galaxies (HAE 16 and HAE 9) using multiwavelength data sets.  We compare the CO-detected galaxies with others in general fields and other protoclusters in terms of CO line widths and luminosities to discuss galaxy evolution in protoclusters.
We summarize the results and discussions in Section~\ref{sec:conclusions}.

Throughout this paper, we assume $H_0 = 67.8$ ${\rm km\,s^{-1}\,Mpc^{-1}}$, $\Omega_0 = 0.308$, and $\Omega_{\Lambda} = 0.692$ \citep{PlanckCollaboration2015}.
The adopted initial mass function (IMF) is the Chabrier IMF in the mass range of $0.1-100\, M_{\odot}$.

\begin{figure*}
\centering
\includegraphics[width=14 cm, bb=0 0 1000 700]{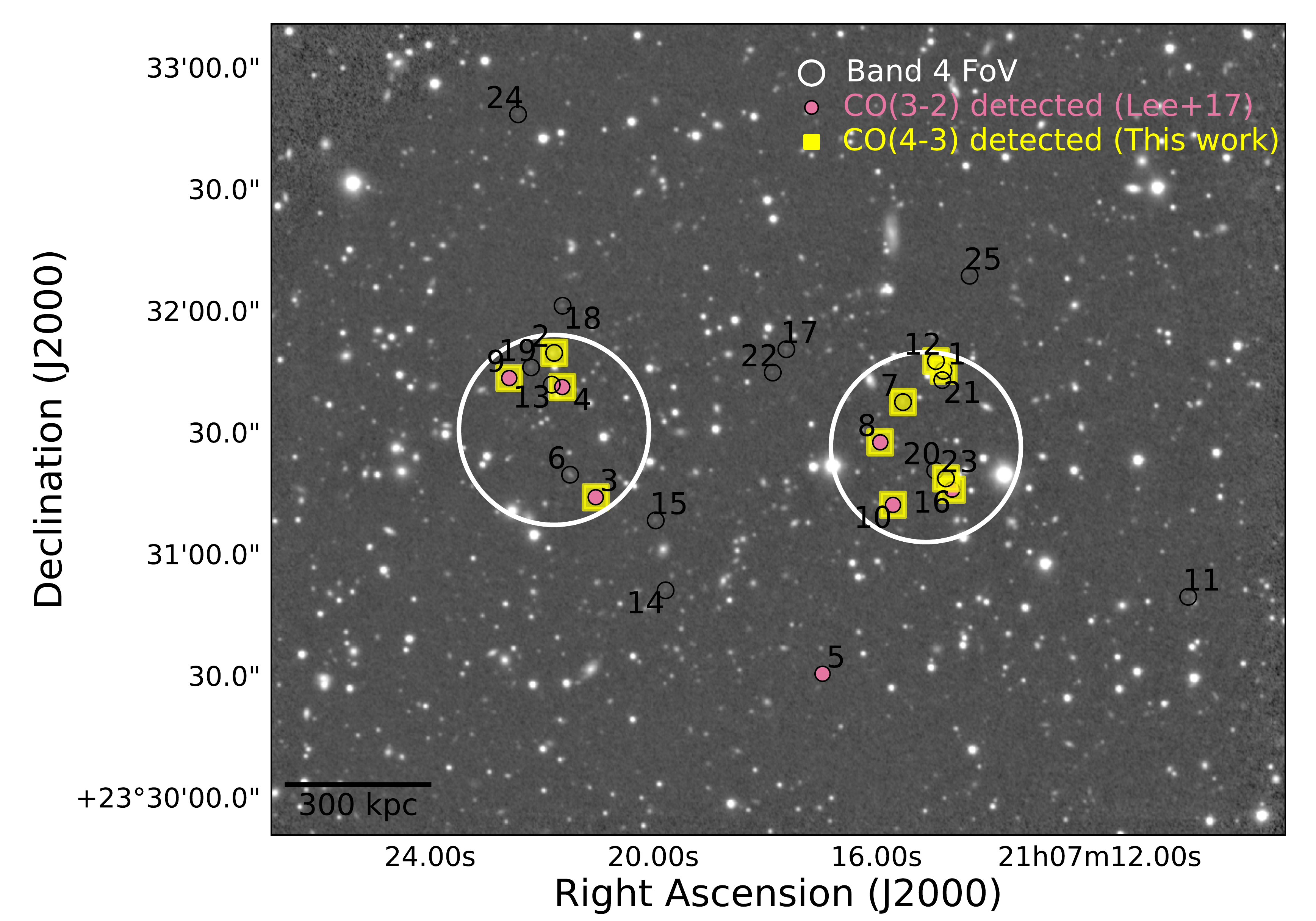}
\caption{The coverage of the Band 4 CO~(4--3) observations shown in white open circles, each corresponding to the ALMA primary beam. The background image is the $K_p$-band image obtained by Subaru/MOIRCS (\citealt{Tanaka2011}). The distribution of the 25 HAEs is shown in black open circles with their ID. Seven HAEs with CO~(3--2) line detection from our previous study (\citetalias{minju2017a}) are shown in pink filled circles. Eleven CO~(4--3) detections are shown in yellow filled squares. \label{fig:fovb4}}
\end{figure*}

\section{Observations and data processing}\label{sec:obs}
\subsection{ALMA Band 4 observations}
We used ALMA Band~4 receivers to obtain the redshifted CO~(4--3) emission (ID : ADS/JAO.ALMA \#2015.1.00152.S, PI : Minju Lee).
We designed our observations to use two pointing positions with the primary-beam size of a $\approx47''$ diameter circle each, i.e., 50\% of the full response of the antenna (Figure~\ref{fig:fovb4}).
A total of 16 out of 25 parent HAEs were targeted in this program.
The field coverage was set differently from the previous CO~(3--2) observations considering the observational efficiency. As a result, six galaxies (HAE 5, 14, 15, 18, 17, 22) were not covered by the CO~(4--3) observations.
We highlight this point especially for HAE~5, which was detected in the CO~(3--2) line.

A total of either 38 or 39 antennas were used with the baseline length ($L_{\rm baseline}$) between 15 m and 3.1 km (C40-6 configuration, which is equivalent to C36-(5)/6 in cycle 3). The total on-source time ($T_{\rm integ.}$) was 2.7 hr with the two-pointing observations, thus $\simeq$ 1.4 hr per pointing.
We used four spectral windows (SPWs) with two placed in the upper and the lower sidebands each.
One SPW was set in the Frequency-Division Mode (FDM) to detect the redshifted CO~(4--3) line by having a channel width of 7.82 MHz ($\sim$ 17.7 km s$^{-1}$) and a total bandwidth of 1.875 GHz.
The remaining three SPWs were observed in the Time-Division Mode (TDM) with a 2.0 GHz bandwidth at 15.6 MHz resolution to cover the dust continuum at 2 mm.
Two quasars, J2148+0657 and J2025+3343 were chosen for bandpass and flux calibration, respectively.
The phase calibrator was J2114+2832.

We used $\mathtt{CASA}$ (\citealt{McMullin2007}) version 4.7.0 
and 5.1.2 for the calibration of visibility data and imaging, respectively.
We first processed with the pipeline script provided by the EA-ARC.
Then additional flagging of $T_{\rm sys}$ outliers was applied manually to improve the image qualities after the inspection of the obtained first images and the pipeline logs. 
We made final images using the recalibrated data sets after flagging.

Images were produced by $\mathtt{CASA}$ task $\mathtt{tclean}$ and deconvolved down to the 2$\sigma$ noise level.
Clean masks were created based on the positions of HAEs, using circular regions with a radius of 3$\times$ the beam size.
The typical noise level in the final image is 0.14 mJy beam$^{-1}$ at 80 km s$^{-1}$.
The synthesized beam is $0''.52\times0''.32$ which is a finer beam by a factor of 2, compared to the beam sizes of our previous ALMA observations of the CO~(3--2) line and 1.1 mm continuum in \citetalias{minju2017a} (i.e., $0''.91\times0''.66$ and $0''.78\times0''.68$ for the CO~(3--2) and the 1.1 mm observations, respectively).
All images were analyzed after subtracting the continuum emission using $\mathtt{uvcontsub}$.
Images without continuum subtraction were also investigated to test whether the continuum has been misidentified as a (broad) line detection.
This was intended for cross-checking the detection with large line widths and for justifying the velocity range used in the continuum subtraction.

\subsection{Detection criteria of CO~(4--3)}\label{sec:criteria}
We regard a galaxy as detected by imposing double-step criteria. First, we select candidates if (a) the signal-to-noise (S/N) ratio of the peak of the velocity-integrated intensity ($I_{\rm CO43,peak}$) is equal to or greater than 4.5 ($S/N_{I_{\rm CO43,peak}}\geq 4.5$). 
The noise estimate for the peak line intensity is based on the calculation after taking into account the channel noise and integrating velocity range, following \cite{Hainline2004}. 
We then investigate whether the peak position spectrum satisfies either of the following criteria: (b) a peak flux density ($S_{\rm CO43, peak}) \geq 3.5~\sigma$, or
(c) at least two continuous channels including a maximum peak-flux channel with fluxes $> 2.5~\sigma$, where $\sigma$ is the average channel noise level, estimated from the line-free positions.
These ``two-step" criteria were successful in detecting CO~(3--2) lines in \citetalias{minju2017a}.
 Provided the success of CO~(3--2) line detection, we opt to apply a similar detection analysis for CO~(4--3) as well. 

Column (9) in Table~\ref{tab:detection} shows the set of criteria satisfied for individual galaxies, which is used to determine the CO~(4--3) line detection. 
One can also check whether or not criteria (a) and (b) are satisfied by referring to columns (4) and (5), respectively, in Table~\ref{tab:detection}.

We note that the S/N requirements used here are slightly different from those used in \citetalias{minju2017a}. Compared to \citetalias{minju2017a}, the requirements are loosened by 0.5 (for conditions ``a" and ``b") and 1 (for condition ``c"), considering the smaller beam size and only a little improvement on the point-source sensitivity. 
It is intended to enhance and not to miss the detection for resolved galaxies with lower fluxes. The loosened criteria do not change the total number of CO~(3--2) detections at the same time with matched spectral resolution (Section~\ref{sec:matching}).
 
We tested the data cubes with a spectral solution of 80 km s$^{-1}$, considering typical line widths due to the galaxy rotation.
Hereafter, we use the velocity resolution of 80 km s$^{-1}$ as a default resolution, or otherwise specified.
We searched for a peak position within a circle of radius $1''.0$, centered on the positions of HAEs. 
The radius is a conservative choice considering the S/N in the intensity map and the beam size.

For the first criterion (a), we determined the integral velocity range by exploring both natural-weighted and lower resolution images. 
The lower resolution images were created by giving less weights to longer-baseline visibilities (\textit{uv} tapering) in order to have a synthesized beam size that is comparable to previous CO~(3--2) observations.
We made the tapered images by modifying the $\mathtt{uvtaper}$ argument in the CASA task $\mathtt{tclean}$ with a parameter of $0''.5$ to produce a synthesized beam of $0''.74\times0''.58$.
Hereafter, we call this lower resolution image the ``tapered" image and the natural-weighted image the ``original" image.

The velocity-integration ranges are determined based on the S/N.
We investigated a wide velocity-integration range with a redshift prior set as the central velocity ($v_0=0$) if the galaxy has, for instance, redshift information determined by the CO~(3--2) line detection.
If not, $z=2.485$ is set to the central velocity as an initial value.
We investigated a velocity-integration range as large as 2000 km s$^{-1}$ with a shift in the central velocity of up to $\pm 1800$ km s$^{-1}$ ($v_0\pm$1800 km s$^{-1}$), which is large enough to cover the expected protocluster members.
The final S/Ns for both integrated flux and peak flux were confirmed after visual inspection of the data cubes and velocity-integrated maps.

\subsection{Reanalysis for the CO~(3--2) data}\label{sec:matching}
We reanalyzed our CO~(3--2) images to have a common spectral resolution with the CO~(4--3) images, because we used the spectral resolution of 100 km s$^{-1}$ in \citetalias{minju2017a}.
The details of the calibration are given in \citetalias{minju2017a}. 
New images were created using CASA version 5.1.2 .
 
The same detection criteria used for CO~(4--3) line detection (Section~\ref{sec:criteria}) are applied for the CO~(3--2) images.
The total detection number of the CO~(3--2) line has not changed ($n=7$) with the new spectral resolution and the loosened criteria, giving support to the justification (or the robustness) of the loosened criteria.
%
%
\section{Results}\label{sec:results}
\subsection{Detection of CO~(4--3)}

A total of 11 HAEs are detected in the CO~(4--3) line, out of the 16 targeted.
The number of CO-detected galaxies has grown by a factor of $\sim2$ with the CO~(4--3) observations.
Figure~\ref{fig:gal1} shows the CO~(4--3) intensity maps and the CO~(4--3) spectra obtained from the original images at a resolution of $0''.52\times0''.32$ overlaying the CO~(3--2) spectra in dashed lines.
We list the line intensities, the peak flux values, the redshifts, and the line widths for the CO~(4--3) detection, as well as the 3~$\sigma$ upper limits of the peak fluxes for nondetections, in Table~\ref{tab:detection}.
We used the primary-beam-corrected images for all these measurements. The line intensities are measured by performing a two-dimensional Gaussian fit on their respective maps with a circular aperture of $0''.7$ using CASA $\mathtt{imfit}$. 
The aperture size is determined by investigating the flux growth curves at various aperture sizes. 
With this aperture, the flux starts to flatten and/or the S/N is the highest value.
The uncertainties of the integrated line flux include both the fitting error and the image noise.
The listed line widths and the redshifts are estimated from a one-component Gaussian-model fitting with errors from the fitting and the assumed spectral resolution.

All six CO~(3--2)-detected HAEs \citepalias{minju2017a} that are located within the observed fields are detected in the CO~(4--3) line (see Section~\ref{sec:co43co32}).
This confirms the detections of two CO lines associated with the HAEs and the protocluster.
Table~\ref{tab:detection} and \ref{tab:co43co32} present two kinds of redshifts obtained from the detected CO~(4--3) line, where the former lists the redshifts estimated from the original data cubes, while the latter shows the ones estimated from the tapered images. The comparison between these two measurements are used for later discussions on the merger nature of HAE~9 (see Section~\ref{sec:co43co32} and Section~\ref{sec:mergers}).

In Appendix~\ref{app:newdetection}, we summarize the five newly detected galaxies (HAE~1, 2, 7, 12, and 23) in the CO~(4--3) line without a previous CO~(3--2) line detection.
These galaxies need additional observations in different CO transitional lines or deeper CO~(3--2) integrations to confirm the CO~(4--3) line detection. 
We note that three of these HAEs (HAE~1, 2, 7) have H$\alpha$ redshifts from Subaru grism spectroscopy (\citealt{Tanaka2011}; I. Tanaka et al. in preparation.) and except for HAE~1, the redshifts from CO and H$\alpha$ are consistent within errors (see Appendix~\ref{app:newdetection} for a potential reason for the difference).
\clearpage
\startlongtable
\movetabledown=2.3in
\begin{rotatetable}
\begin{deluxetable*}{ccccccccccc}
\tablecaption{Identification of CO~(4--3) line for HAEs \label{tab:detection}}
\tabletypesize{\footnotesize}
\tablewidth{0.99\textwidth}
\tablehead{
\colhead{ID} & \colhead{R.A.\tablenotemark{$^{\Diamond}$}}  & \colhead{Decl.\tablenotemark{$^{\Diamond}$}} 	&	\colhead{$I_{\rm CO43, peak}$}	&	\colhead{$S_{\rm peak,CO43}$}	&	\colhead{$I_{\rm CO43, gaussian}$} 	&  \colhead{$z_{\rm CO43}$} & \colhead{FWHM} & \colhead{Criteria\tablenotemark{$\dagger$}}  & \colhead{CO~(4--3)}	& \colhead{CO~(3--2) \tablenotemark{$\ddagger$}}\\
& \colhead{(J2000)}  & \colhead{(J2000)} & \colhead{(Jy km s$^{-1}$ beam$^{-1}$)}& \colhead{(mJy beam$^{-1}$)} & \colhead{(Jy km s$^{-1}$)} &  & \colhead{(km s$^{-1}$)} & \colhead{ (CO43/CO32)}  & \colhead{detection?}	& \colhead{detection? }\\
\colhead{(1)} & \colhead{(2)}  & \colhead{(3)} &\colhead{(4)} & \colhead{(5)} & \colhead{(6)} &  \colhead{(7)} & \colhead{(8)} & \colhead{(9)}  & \colhead{(10)}	& \colhead{(11)}}
\startdata
HAE~1 &    21:07:14.802 &+23:31:44.593& 0.32$\pm$0.06& 1.06$\pm$0.25    & 0.45$\pm$0.13 & 2.4945$\pm$0.0009  & 414$\pm$123 & abc/none   &y     & n\\ 
HAE~2 &    21:07:21.739 &+23:31:50.123& 0.20$\pm$0.04& 0.79$\pm$0.22    & 0.13$\pm$0.05 & 2.4893$\pm$0.0009  & 267$\pm$114 &ab/none   &y     & n\\ 
HAE~3 &    21:07:21.030 &+23:31:13.922& 0.36$\pm$0.05& 0.72$\pm$0.20    & 0.48$\pm$0.12 & 2.4889$\pm$0.0009  & 456$\pm$126 &abc/ac   &y     & y\\ 
HAE~4 &    21:07:21.633 &+23:31:41.273& 0.12$\pm$0.02& 0.65$\pm$0.17    & 0.15$\pm$0.05 & 2.4916$\pm$0.0009  & 199$\pm$96 &abc/abc   &y     & y\\ 
HAE~6&    21:07:21.492 &+23:31:19.524& -& $<$0.54   & - & - & -&none/none     &n      & n\\ 
HAE~7 &    21:07:15.522 &+23:31:37.543& 0.47$\pm$0.06& 0.75$\pm$0.18    & 0.48$\pm$0.12 & 2.4825$\pm$0.0010  &966$\pm$206 &ab/none   &y     & n\\ 
HAE~8 &    21:07:15.943 &+23:31:27.742& 0.18$\pm$0.02& 1.18$\pm$0.18    & 0.45$\pm$0.07 & 2.4855$\pm$0.0009  & 112$\pm$91 &abc/abc   &y     & y\\ 
HAE~9 &    21:07:22.586 &+23:31:43.272& 0.35$\pm$0.05& 0.94$\pm$0.22    & 0.63$\pm$0.16 & 2.4830$\pm$0.0009  & 469$\pm$126 &abc/abc   &y     & y\\ 
HAE~10 &    21:07:15.725 &+23:31:12.142& 0.21$\pm$0.04& 0.58$\pm$0.21    & 0.30$\pm$0.11& 2.4889$\pm$0.0010  & 614$\pm$201 &ac/ac   &y     & y\\ 
HAE~12 &    21:07:14.904 &+23:31:48.193& 0.18$\pm$0.04& 1.05$\pm$0.28    & 0.20$\pm$0.07& 2.4937$\pm$0.0009  & 171$\pm$99 &abc/none   &y     & n\\ 
HAE~13&    21:07:21.820 &+23:31:41.747& - & $<$0.53   & -& - & -& a/none       &n & n\\ 
HAE~16 &    21:07:14.642 &+23:31:15.593& 0.31$\pm$0.03& 0.90$\pm$0.18    & 0.53$\pm$0.10 & 2.4848$\pm$0.0009  & 348$\pm$98 &abc/abc   &y     & y\\ 
HAE~19&    21:07:22.192 &+23:31:45.995& - & $<$0.63  & - & - & -&none/none    &n & n\\ 
HAE~20&    21:07:14.946 &+23:31:20.572& -& $<$0.47   & - &- & -&none/none   &n   & n\\ 
HAE~21&    21:07:14.820 &+23:31:42.856& -& $<$0.66   & - &- & -&none/none   &n   & n\\ 
HAE~23 &    21:07:14.773 &+23:31:18.443& 0.23$\pm$0.04& 0.58$\pm$0.16	& 0.23$\pm$0.06 & 2.4774$\pm$0.0010 & 631$\pm$203 & ab/none   &y     & n\\\hline
\enddata
\tablecomments{This table lists the CO~(4--3) line information analyzed in the map and spectrum with the original beam size of $0''.52\times0''.32$ and at 80 km s$^{-1}$ resolution, respectively. Columns : (1) HAE ID; (2) right ascension (R.A.) for the peak position of either CO~(4--3) (if detected) or H$\alpha$ emissions (if not); (3) declination (decl.) for the peak position of either CO~(4--3) (if detected) or H$\alpha$ emissions (if not); (4) peak intensity of CO~(4--3); (5) peak flux of CO~(4--3); (6) CO~(4--3) line intensity based on a Gaussian fit with an aperture of $0''.7$ using CASA $\mathtt{imfit}$ and the errors include both the fitting errors and image noise; (7) the redshift estimated from the 1D Gaussian fit of the CO~(4--3) spectrum, and the errors are fitting errors including the assumed spectral resolution, i.e., 80 km s$^{-1}$; (8) the estimated line width with fitting errors; (9) detection criteria that are satisfied with the same (loosened) criteria used for the CO~(4--3) and CO~(3--2) detections; (10) information on whether the CO~(4--3) line is detected (y=yes) or not (n=no); (11) information on whether the CO~(3--2) line is detected (y=yes) or not (n=no). $\Diamond$: We use the position of CO~(4--3) if detected in CO~(4--3) line. Otherwise, we adopt the position of H$\alpha$ from Subaru/MOIRCS observations (\citealt{Tanaka2011}; I. Tanaka et al. 2019, in preparation), $\dagger$: Detection criteria (a) the peak intensity (column 4) S/N $> 4.5$ ; (b) peak flux (column 5) $> 3.5~\sigma$; (c) at least two continuous channels including a maximum peak flux with S/N $> 2.5~\sigma$, $\ddagger$: CO~(3--2) detection using the same criteria applied within the paper, reconfirming the previous detection of HAE~3, 4, 8, 9, 10, 16 presented in \citetalias{minju2017a}, at matched 80 km s$^{-1}$ resolution with CO~(4--3).}
\end{deluxetable*}
\end{rotatetable}
\clearpage
\startlongtable
\movetabledown=1.25in
\begin{rotatetable}
\begin{deluxetable*}{ccccccccc}
\tablecaption{Summary of the CO~(4--3) and CO~(3--2) emissions matched for spatial and spectral resolutions\label{tab:co43co32}}
\tabletypesize{\footnotesize}
\tablewidth{0.99\textwidth}
\tablehead{
\colhead{ID} &\colhead{$I_{\rm CO43, peak, tapered}$}  & \colhead{$I_{\rm CO32, peak}$} & \colhead{$S_{\rm peak,CO43, tapered}$}  &  \colhead{$S_{\rm peak,CO32}$} &\colhead{$z_{\rm CO43, tapered}$}  & \colhead{$z_{\rm CO32}$} & \colhead{FWHM$_{\rm CO43}$} &  \colhead{FWHM$_{\rm CO32}$} \\
\colhead{} &  \colhead{(Jy km s$^{-1}$ beam$^{-1}$)}& \colhead{(Jy km s$^{-1}$ beam$^{-1}$)}& \colhead{(mJy beam$^{-1}$)}  &  \colhead{(mJy beam$^{-1}$)} & &  &  \colhead{km s$^{-1}$} &   \colhead{km s$^{-1}$} }
\startdata
HAE~3 &  0.43$\pm$0.06& 0.36$\pm$0.06    & 0.64$\pm$0.30 &0.88$\pm$0.27& 2.4889$\pm$0.0009 &  2.4890$\pm$0.0009  & 471$\pm$158& 459$\pm$129\\ 
HAE~4 &  0.13$\pm$0.03& 0.21$\pm$0.03    & 0.42$\pm$0.22 &0.80$\pm$0.20& 2.4913$\pm$0.0009 & 2.4899$\pm$0.0009	& 211$\pm$110 &299$\pm$112\\ 
HAE~8 &  0.27$\pm$0.02& 0.29$\pm$0.03    & 1.13$\pm$0.22 &1.27$\pm$0.21& 2.4857$\pm$0.0009 &  2.4865$\pm$0.0009& 153$\pm$83 &193$\pm$88 \\ 
HAE~9 &  0.65$\pm$0.08& 0.51$\pm$0.05    & 0.78$\pm$0.26 &1.13$\pm$0.21& 2.4841$\pm$0.0009 & 2.4858$\pm$0.0009& 800$\pm$165 &732$\pm$119\\ 
HAE~10 &  0.25$\pm$0.05& 0.35$\pm$0.05    & 0.46$\pm$0.26 &0.90$\pm$0.28& 2.4879$\pm$0.0010 & 2.4860$\pm$0.0009& 610$\pm$237& 357$\pm$117\\ 
HAE~16 &  0.40$\pm$0.04& 0.52$\pm$0.07    & 0.69$\pm$0.21 &1.07$\pm$0.31& 2.4842$\pm$0.0009 & 2.4846$\pm$0.0009  & 452$\pm$101 &547$\pm$138 \\ \hline
\enddata
 \tablecomments{We compare the CO~(4--3) map with that of CO~(3--2) after matching the spatial resolution of CO~(4--3) to that of CO~(3--2) by making low spatial resolution data for CO~(4--3). The resolution achieved by $uv$ tapering is $0''.74 \times 0''.58$ for $0''.52\times 0''.32$, compared to $0^{\prime\prime}.91\times0^{\prime\prime}.66$ for CO~(3--2). Strictly speaking, this is slightly smaller than that of CO~(3--2), but we chose this resolution for comparison considering the noise level (or S/N), instead of a larger beam that can be obtained by weighting more on shorter baseline data sets, which give smaller S/N. The CO~(3--2) data presented in \citetalias{minju2017a} have been reanlayzed here, using the same detection criteria applied within this paper by matching the spectral resolution at 80 km s$^{-1}$.}
\end{deluxetable*}
\end{rotatetable}
\clearpage

\subsection{Consistency between CO~(4--3) and CO~(3--2)}\label{sec:co43co32}
All CO~(3--2)-detected HAEs other than HAE~5, which was outside the fields of view (FoVs; Figure~\ref{fig:fovb4}), are detected in the CO~(4--3) line.
We matched the angular resolution and the spectral resolution, as explained in Section~\ref{sec:criteria} and Section~\ref{sec:matching} to investigate the consistency.
Table~\ref{tab:co43co32} lists the line information after the matching processes\footnote{In \citetalias{minju2017a}, we listed the CO~(3--2) redshifts based on the median frequencies used for integration instead of the Gaussian-model fit considering the low S/N and the broad nature of several galaxies. Here, we enforced the Gaussian fitting on the CO~(3--2) spectrum for uniform analysis.}.

\begin{figure*}
\centering
\includegraphics[width=15.5cm, bb=0 0 1100 800]{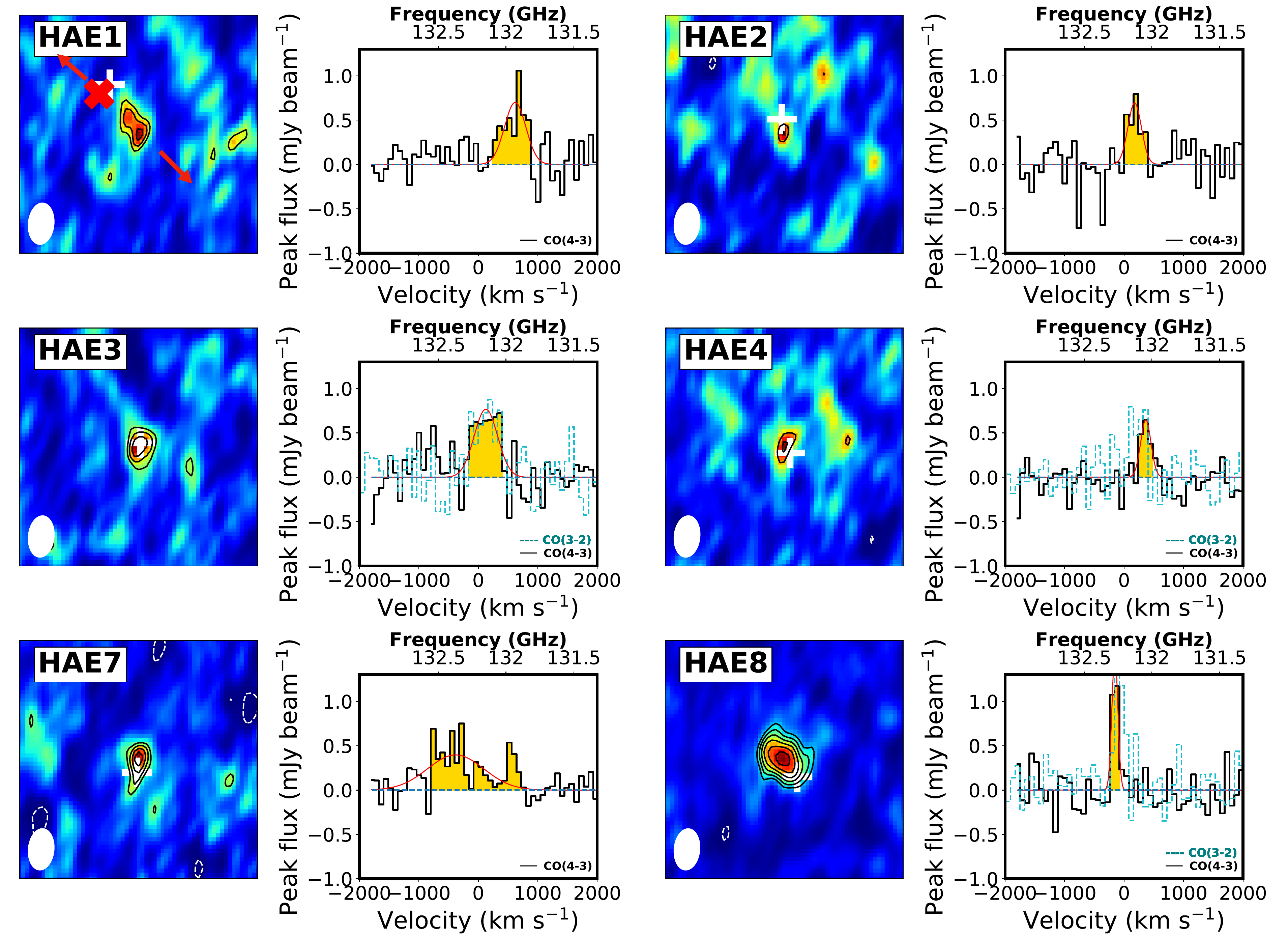}
\caption{A gallery of HAEs with CO~(4--3) line detection. We show the intensity map on the left and spectrum on the right for individual galaxies. The panel size of the intensity map is $3''$ in width. The beam size of $0''.52\times0''.32$ is shown as a white filled ellipse. The contour level is shown in steps of 1$\sigma$ starting from 4$\sigma$, i.e., 4$\sigma$, 5$\sigma$, 6$\sigma$, ... . We also plot the negative 4$\sigma$ contour in white dashed lines. We plot the peak positions of the H$\alpha$ emission in white crosses. For HAE~1, we indicate the peak position of the radio continuum at 3~GHz (S band) and 6~GHz (C band) with a red cross (``${\bf \times}$'') and the direction of its bipolar radio jet with red arrows. The spectrum is shown in units of mJy beam$^{-1}$ measured at the peak position with a spectral resolution of 80 km s$^{-1}$. We fit the spectrum with a one-dimensional, one-component Gaussian model, which is shown in red solid lines. The reference velocity ($v=0$) is based on the mean redshift of the total CO~(4--3) detection, which is $z_{\rm mean, 4C~23.56}=2.4872$. For HAE~3, 4, 8, 9, 10, and 16, the CO~(3--2) spectra are overlaid in teal dashed lines. The CO~(3--2) spectra shown here are drawn from the peak position with a synthesized beam of $0''.91 \times 0''66$. \label{fig:gal1}}
\end{figure*}
\setcounter{figure}{1}  
\begin{figure*}
\centering
\includegraphics[width=15.5cm, bb=0 0 1100 800]{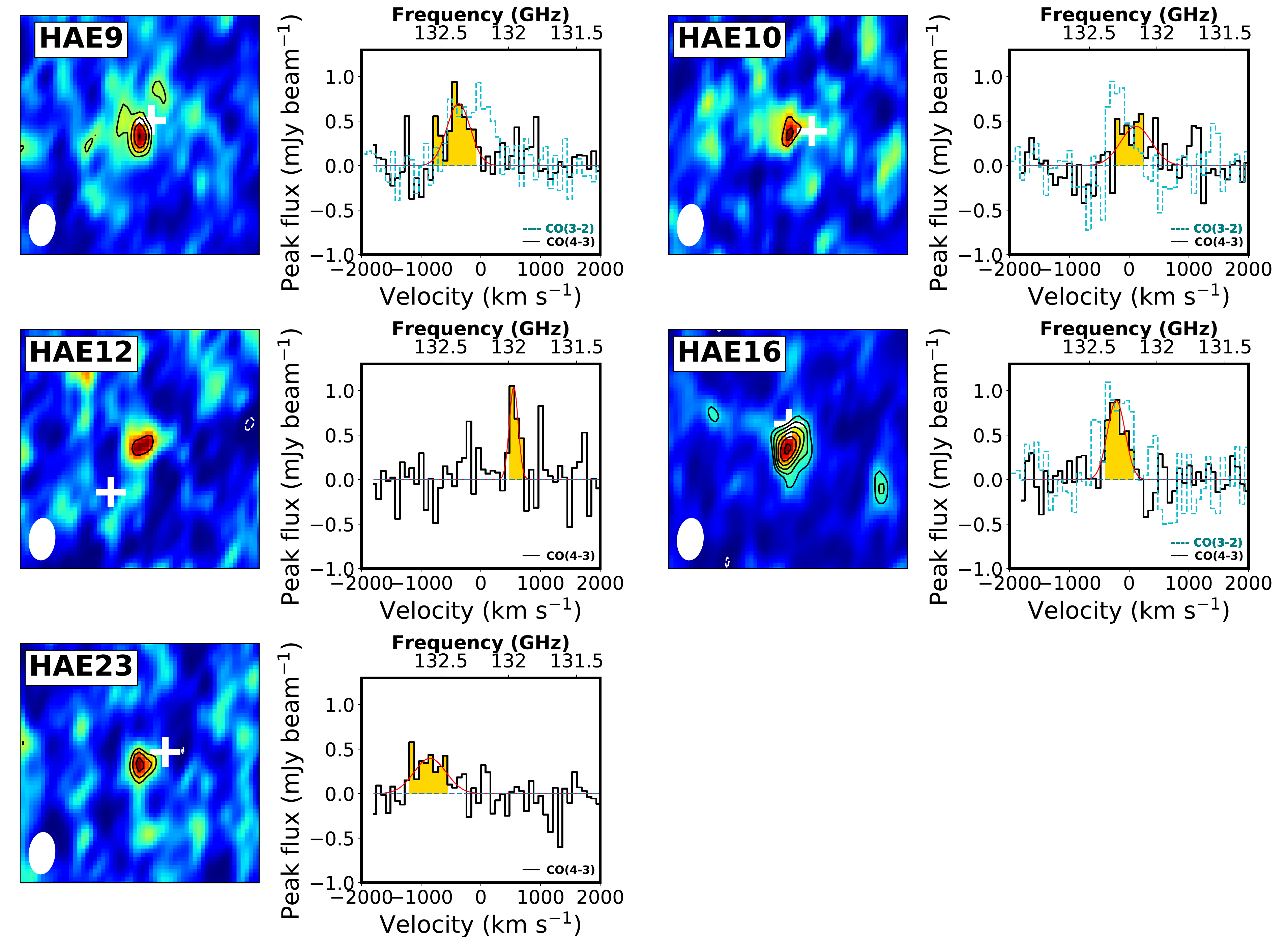}
\caption{(continued) A gallery of HAEs with CO~(4--3) line detection. \label{fig:gal2}}
\end{figure*}

The spectroscopic redshifts derived independently from the CO~(4--3) (tapered) and CO~(3--2) data cubes are consistent within $|\Delta z_{43-32}| = |z_{\rm CO43, tapered}-z_{\rm CO32, original}| \leq 0.0017$, corresponding to the velocity difference of $\pm150$~km s$^{-1}$\deleted{at $z\simeq2.5$}.
The FWHM values are also consistent within the errors.
Therefore, we conclude that both CO lines are simultaneously detected and consistent with respect to the line redshifts and the line widths.

Unlike  other galaxies, for HAE~9, we find that the redshift defined by CO~(3--2) is not consistent with the value obtained from the original map of CO (4--3), which is indicated in Figure~\ref{fig:gal2} visually.
The spectrum of the CO~(3--2) line is shown as a teal dashed line, which is inconsistent with that of the CO~(4--3) line, shown as a black solid line.
The redshift is different by $|z_{\rm CO43, natural} - z_{\rm CO32, natural}| = 0.0028$.
We explain the details of this discrepancy in Section~\ref{sec:subcomponent} and discuss the nature of HAE~9 in Section~\ref{sec:mergers}.

The positional differences ($\Delta$) between the peaks of different CO lines are within $0''.3$ for most cases. 
We conclude that the CO~(4--3) and CO~(3--2) line emissions are coming from the same region, at least globally. 
The positional differences are all consistent within the positional errors considering the S/Ns and the synthesized beam of the tapered map.

One exception is HAE~4 ($\Delta \approx 0''.6$).
In HAE~4, the peak position of the CO~(4--3) emission is closer to the peak of H$\alpha$ (NB) emission (Figure~\ref{fig:gal1}) than that of the CO~(3--2) emission.
As discussed in \citetalias{minju2017a}, the H$\alpha$ emission shows an extended structure, which we attributed the peak differences to the internal structure of the galaxy and/or dust extinction.
The difference in the peak position between CO~(4--3) and CO~(3--2) might have originated from the different excitation conditions within the galaxy structure.
The origin of the positional difference between the line emissions needs to be confirmed by deeper, higher angular resolution observations to understand the nature of this massive galaxy.

Finally, the observed line intensity ratios between CO~(4--3) and CO~(3--2), $I_{\rm co43}/I_{\rm co32}$ are spread around unity, i.e., $I_{\rm co43}/I_{\rm co32} = 0.63-1.26$ with a median of 0.86.
The median is similar to the typical line ratio observed for high-$z$ normal disk-like galaxies (e.g., \citealt{Daddi2015}) or local LIRGs, which is slightly lower than that observed in local ULIRGs (e.g.  \citealt{Papadopoulos2012}).
In this paper, we defer the discussions on CO spectral line energy distributions (SLEDs) and gas excitation for the protocluster galaxies because these cannot be done with only two CO detections in close transition ($J=4-3$ and $J=3-2$).
%
\subsection{Subcomponents of gas-rich galaxies}\label{sec:subcomponent}
\begin{figure*}
\centering
\includegraphics[width=0.7\textwidth, bb=0 0 1200 1000]{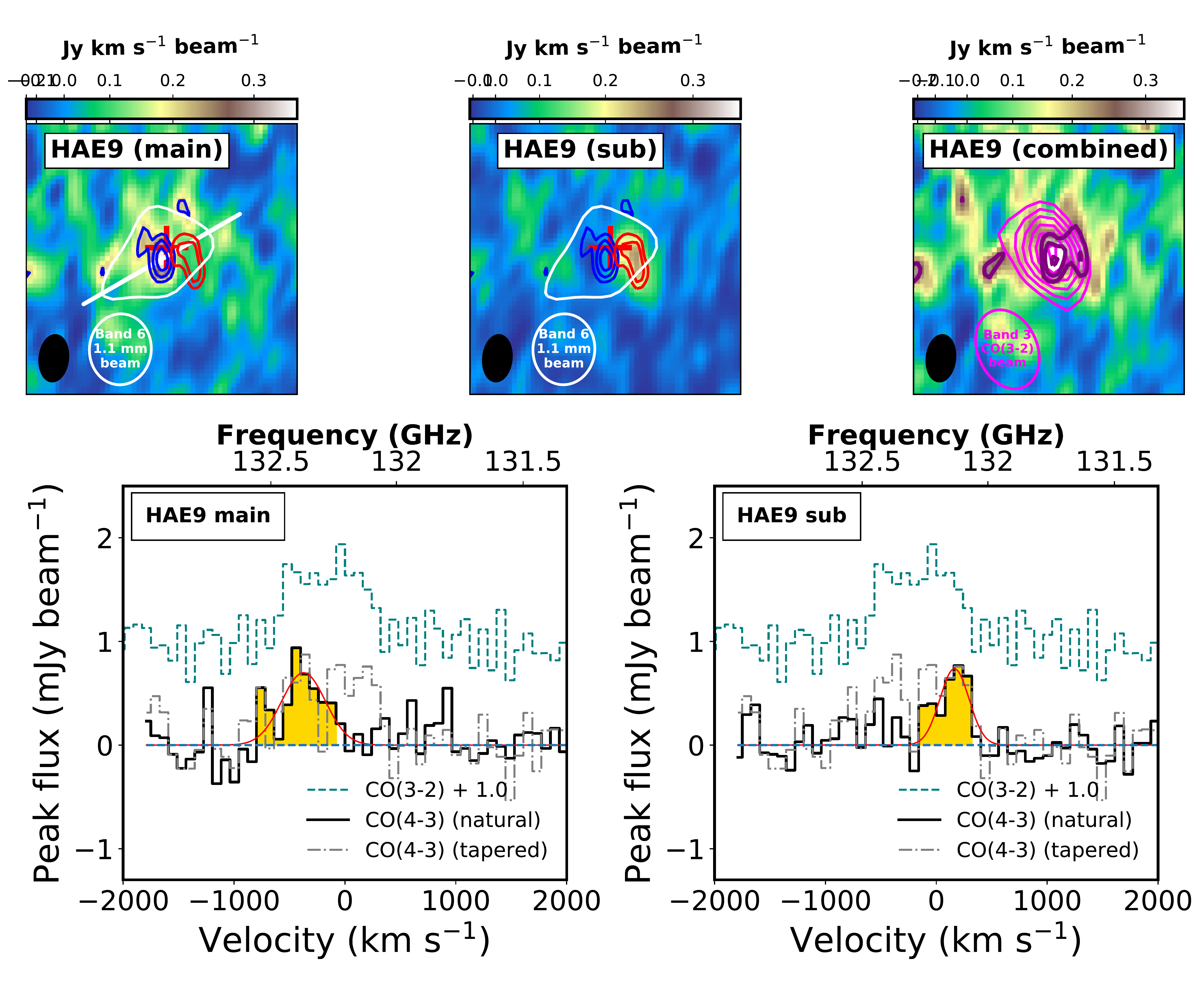}
\caption{Subcomponents of HAE~9 suggesting a gas-rich merger. Top left : the color image for the main component of HAE~9 with a peak $S/N$ of $\sim7$. A white line running in the direction of northwest to southeast is the (morphological) position angle of the 1.1 mm dust continuum obtained from $\mathtt{uvmodelfit}$ in CASA (see Section~\ref{sec:mergers}).
Top middle : the color image of the subcomponent of HAE~9, which is redshifted by $\sim$ 500 km s$^{-1}$ from the main component.
We show the main and the subcomponent in blue and red contours, respectively. 
The single white contour corresponds to the $4\,\sigma$ signal of the 1.1 mm dust continuum. The red cross is the peak position of the CO~(3--2) emission.
Top right: the CO~(4--3) intensity map, which is obtained by integrating the entire channel ranges of the main and the subcomponents, is shown in purple contours. Contours in magenta show the previous CO~(3--2) detection for comparison. All contours in the upper panels are in steps of $1~\sigma$, starting from $4~\sigma$, i.e., 4$\sigma$, 5$\sigma$, 6$\sigma$, ..., .
The black filled ellipse is the beam size of the CO~(4--3) observations (this work), while the white open ellipse is the beam size of 1.1 mm dust continuum in ALMA Band~6 at 1.1 mm and the magenta open ellipse is the CO~(3--2) beam size (\citetalias{minju2017a}).
Bottom panels: the CO~(4--3) spectrum of the main component (left) and the subcomponent (right). The CO~(4--3) spectrum from the tapered image and the CO~(3--2) spectrum are also plotted with the gray dashed-dotted line and the teal dashed line, respectively. For the CO~(3--2) spectrum, we shifted the baseline to 1.0 for better visual inspection. \label{fig:merger_hae9}}
\end{figure*}

\begin{figure*}
\centering
\includegraphics[width=0.7\textwidth, bb=0 0 1200 1000]{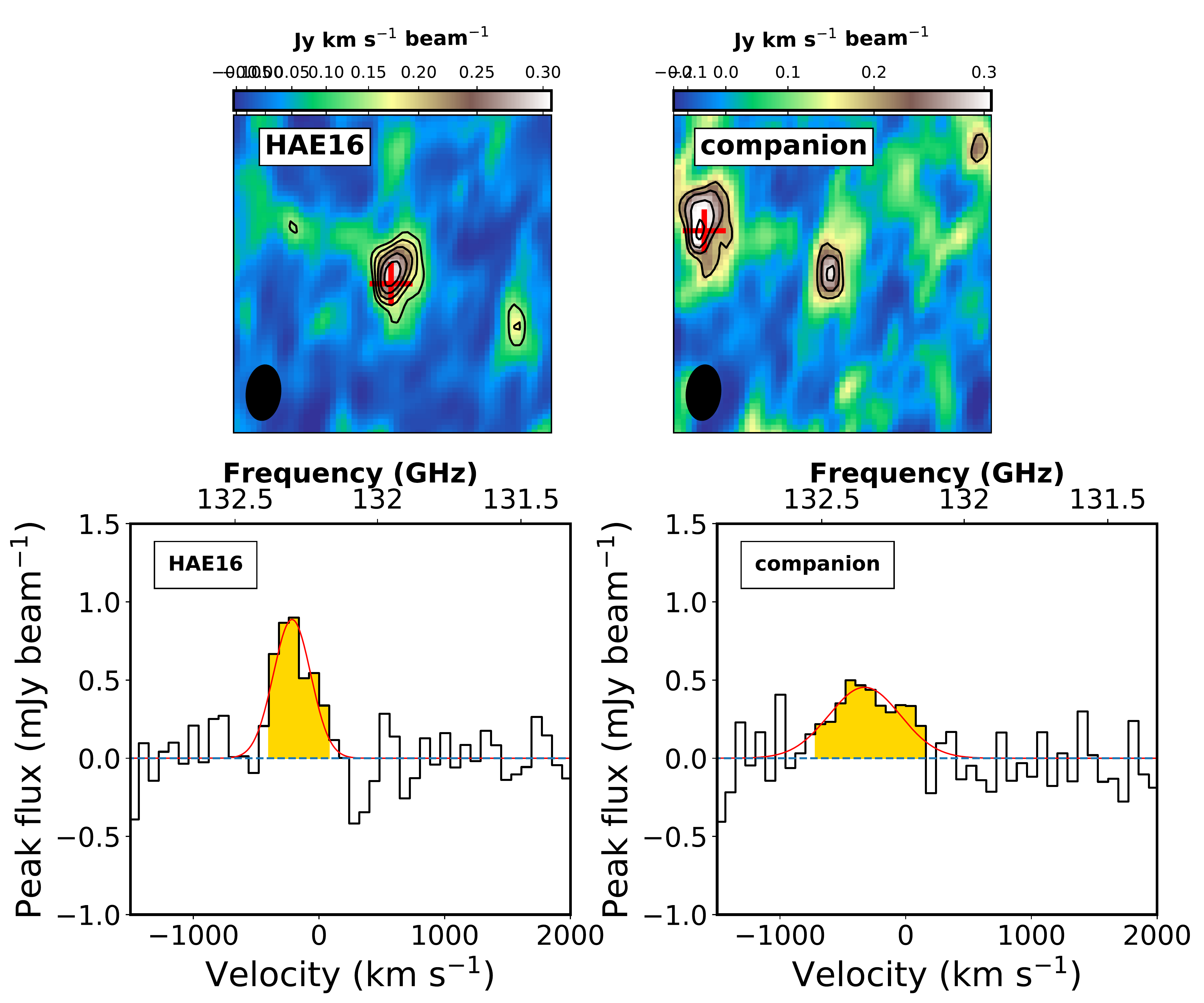}
\caption{HAE~16 and its potential minor merging galaxy. Top left: the HAE~16 CO~(4--3) intensity map. Top right : potential minor merger that can be ``accreted" to HAE~16, which is blue-shifted by $\sim$ 100 km s$^{-1}$.
Contours are shown in steps of $1~\sigma$ starting from $4~\sigma$. 
The red cross is the peak position of the CO~(3--2) emission.
Bottom panels: the CO~(4--3) spectrum of HAE~16 (left) and the companion (right) for comparison. \label{fig:merger_hae16}}
\end{figure*}

We highlight two gas-rich ($f_{\rm gas}>0.7$) galaxies detected at a relatively high significance, HAE~9 and HAE~16.
We find evidence of gas-rich mergers by exploring the data cubes with different integration ranges around the positions of the HAEs.

For HAE~9, we confirmed two distinct components which are spatially and spectroscopically resolved.
The CO~(3--2) observations, which have  coarser spatial resolution, could not resolve this galaxy.
Given the discrepancies between the CO~(3--2) and the CO~(4--3) lines in the original data cubes, we investigated the CO~(4--3) cubes again. 
An additional component just next to the ``main" component that still satisfies our detection criteria is identified. 
Related to this, we emphasize that our detection procedure is optimized to search for a peak and a channel range that delivers the largest S/N.
As such, additional ``sub"components at lower significance, if any, would be missed.
The blue-shifted component of CO~(3--2) has been identified as a ``main component" of CO~(4--3) in the first run of the detection analysis.
By comparing the CO~(4--3) spectrum with that of CO~(3--2) and searching for the red-shifted component in the image, we could confirm another distinct, spatially offset component in the western direction, which still satisfies our detection criteria with the peak intensity $S/N_{I_{\rm CO43,peak}}$ of 6.4 but was missed in our first detection run.
We name this a subcomponent of HAE~9 hereafter.
We show the main and the subcomponents and the spectra from tapered/original CO~(4--3) cubes as well as from CO~(3--2) in Figure~\ref{fig:merger_hae9}.

Another galaxy, HAE~16, has a potential minor merging subcomponent, which appears to be gas rich given the detection of CO~(4--3).
The spectrum of the companion is shown in Figure~\ref{fig:merger_hae16}.
This potential minor merging companion is found southwest of HAE~16, which is detected with $S/N_{I_{\rm CO43,peak}}$ of $\approx 6$ and with a spatial separation of $1''.3$, corresponding to $\sim 10$~kpc. The redshift difference between HAE~16 and the subcomponent is $\Delta z = 0.0013$ (i.e., velocity offset $\simeq 100$~km s$^{-1}$).

We investigated the chances of false-positive detection using $\mathtt{astrodendro}$\footnote{\url{http://www.dendrograms.org/}}.
We ran this with positive and inverse (negative) S/N maps obtained by dividing the original data cubes and the inverse cubes (original data cube $\times (-1)$) with individual channel noises, respectively.
We tested both cubes with searching parameters of $\mathtt{min\_value}=2.5$ (i.e., minimum peak S/N), $\mathtt{min\_delta}=1.0$ (i.e., the significance of the identified ``clump" above $\mathtt{min\_value}$), and $\mathtt{min\_npix}=10$ (i.e., the minimum number of pixels = 10) with a pixel size of $0''.05\times0''.05$, which are set similarly to our detection criteria as much as possible.
We found that the number of line candidates with $S/N_{I_{\rm CO43,peak}} \geq 6.0$ with a line width $\geq 160$ km s$^{-1}$, is seven for the positive cube, while the total number with the same conditions in the negative cube is one; thus, the chance of a false positive is $\approx14\%(=1/7)$ for this field with the given S/N threshold.
Although this kind of blind line-detecting analysis is not the same as our two-step detection criteria and needs future confirmation, we conclude it is less likely that the minor merging companion for HAE~16 is a false-positive detection.
 
\section{Halo-mass estimate of the protocluster}\label{sec:halos}
\begin{figure*}
\centering
\includegraphics[width=0.9\textwidth, bb = 0 0 1000 900]{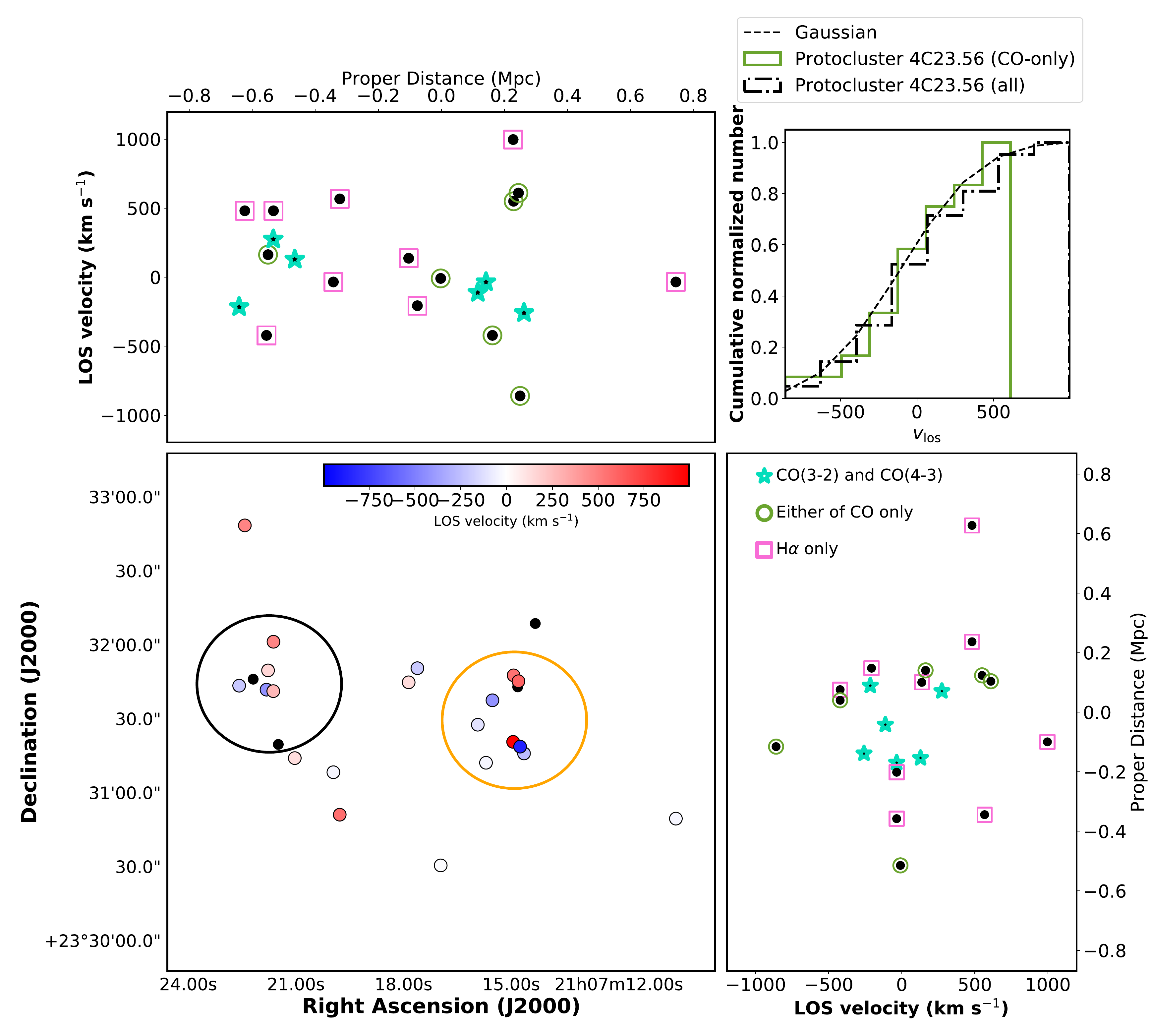}
\caption{Bottom left: the distribution of HAEs shown by the small filled circles. We gathered redshift information from CO (\citetalias{minju2017a} and this paper) and H$\alpha$ (\citealt{Tanaka2011}; I. Tanaka et al. 2019, in preparation) for this plot (Table~\ref{tab:redshifts}). We filled the LOS velocity colors where the scale is indicated by the color bar if the redshift is estimated from CO detection and/or the H$\alpha$ grism; otherwise, they are in black. 
The protocluster center is defined by using the biweight average, which is $\bar{z}_{\rm proto, CO}=2.4874$ (see Section~\ref{sec:halomass} for more details). The large black (the East Clump; EC) and yellow (the West Clump; WC) circles represent the core radius (0.8 comoving Mpc) of a protocluster that will become a cluster with a halo mass of $10^{14}~M_{\odot}$, as expected in \citet[Section~\ref{sec:subhalo}]{Chiang2017}. Top left: the velocity--position distributions projected on the R.A. axis. Bottom right : the velocity--position distributions projected on the decl. axis. Symbols of stars, circles, and squares are used to indicate simultaneous CO~(4-3) and CO~(3--2) detection, one CO-line, and H$\alpha$-only detection cases, respectively. Top right: the cumulative velocity distribution for the normalized galaxy number. For reference, we also plot a cumulative profile of a Gaussian velocity distribution using the derived velocity dispersion ($\sigma_{\rm BI} = 421$~km s$^{-1}$) of the 4C~23.56 protocluster.\label{fig:3ddist}} 
\end{figure*}

\subsection{Overall halo estimates}\label{sec:halomass}
A halo-mass estimate for protoclusters is valuable to test whether or not the protocluster can evolve into a present-day massive cluster.
Under the assumption of virialization, velocity dispersion offers a rough measure of the halo mass of the protocluster. 
We estimate the velocity dispersion based on the velocity differences with respect to the protocluster redshift. 

To define the protocluster redshift, we combined the redshift information from our CO~(4--3) and CO~(3--2) observations, where a total of 12 CO redshifts are available.
We used the mean value of the redshifts listed in Table~\ref{tab:co43co32} for galaxies with simultaneous CO detection.
It is validated by small differences between the redshifts estimated individually from two CO lines (see Section~\ref{sec:co43co32}). 
HAE~9 is a special case in that we used the main (the brightest or secure) component listed in Table~\ref{tab:detection} instead of taking the mean (see Section~\ref{sec:subcomponent} and \ref{sec:mergers}).
The discussion of the halo mass does not change even if we add the subcomponent of HAE~9 for the order estimate. 

We list the redshift information in Table~\ref{tab:redshifts} that we used for the estimate of the halo mass. 
In the table, we also included the H$\alpha$ redshifts if available from the Subaru grism observations (\citealt{Tanaka2011}; I. Tanaka et al. 2019 in preparation). 
The redshift uncertainty for H$\alpha$ is several factors larger than the CO measurements, corresponding to several hundred km s$^{-1}$.

\begin{table}[htp]
\caption{Redshifts used for protocluster halo-mass estimation}
\begin{center}
\begin{tabular}{ccc}
\hline\hline
HAE ID & Redshift	& Lines \\
\hline
HAE~1\tablenotemark{a} & $2.4945\pm 0.0009$ & CO~(4--3)\\ 
HAE~2\tablenotemark{b} & $2.4893\pm 0.0009$	& CO~(4--3)\\ 
HAE~3\tablenotemark{b} & $2.4889\pm 0.0009$	& CO~(4--3), CO~(3--2)\\ 
HAE~4\tablenotemark{b} & $2.4906\pm 0.0009$	& CO~(4--3), CO~(3--2)\\ 
HAE~5\tablenotemark{b} & $2.4873\pm 0.0009$	& CO~(3--2)\\ 
HAE~7\tablenotemark{b} & $2.4825\pm 0.0010$	& CO~(4--3)\\ 
HAE~8\tablenotemark{b} & $2.4861\pm 0.0009$	& CO~(4--3), CO~(3--2)\\ 
HAE~9 & $2.4849\pm 0.0009$ & CO~(4--3), CO~(3--2)\\ 
HAE~10\tablenotemark{b} & $2.4870\pm 0.0010$	& CO~(4--3), CO~(3--2)\\ 
HAE~11 & $2.4870\pm 0.0070$ & H$\alpha$\\ 
HAE~12 & $2.4938\pm 0.0009$ & CO~(4--3)\\ 
HAE~13 & $2.4825\pm 0.0070$ & H$\alpha$\\ 
HAE~14 & $2.4940\pm 0.0070$ & H$\alpha$\\ 
HAE~15 & $2.4870\pm 0.0070$ & H$\alpha$\\ 
HAE~16 & $2.4844\pm 0.0009$ & CO~(4--3), CO~(3--2)\\ 
HAE~17 & $2.4850\pm 0.0070$ & H$\alpha$\\ 
HAE~18 & $2.4930\pm 0.0070$ & H$\alpha$\\ 
HAE~20 & $2.4990\pm 0.0070$ & H$\alpha$\\ 
HAE~22 & $2.4890\pm 0.0070$ & H$\alpha$\\ 
HAE~23 & $2.4774\pm 0.0010$ & CO~(4--3)\\ 
HAE~24 & $2.4930\pm 0.0070$ & H$\alpha$\\ 
\end{tabular}
\end{center}
\tablecomments{\tablenotemark{a} : Redshifted with respect to H$\alpha$ emission (see Appendix~\ref{app:newdetection}).  \tablenotemark{b} : detected also in H$\alpha$ (I. Tanaka et al. 2019, in preparation) and consistent within errors of H$\alpha$ spectroscopy with $R=500$ (i.e., $\Delta z = 0.0070$). We note that the redshift errors for CO lines are the systematic error coming from the assumed spectral resolution of 80 km s$^{-1}$.}
\label{tab:redshifts}
\end{table}%

First, we define the protocluster redshift and this will be used as $v=0$, or the systemic velocity of the protocluster.
We use the (one-step) biweight average (\citealt{Beers1990}) of the member galaxy redshifts ($\bar{z}_{\rm BI}$). 
The biweight mean is more resistant (i.e., hardly changeable) in the presence of outliers compared to the simple mean, assuming that the observed radial velocities are drawn from the Gaussian parent population.
With CO detections, the protocluster redshift is defined as $\bar{z}_{\rm BI, CO} = 2.4874$. For comparison, if the H$\alpha$ redshifts in Table~\ref{tab:redshifts} are included, the protocluster redshift is $\bar{z}_{\rm BI, all} = 2.4880$.
 Hereafter, we use the protocluster redshift obtained from the CO measurements, i.e., $z_{\rm proto} =\bar{z}_{\rm BI, CO}$ for the estimate of the halo mass.

Then, a line-of-sight (LOS) proper (or peculiar) velocity $v_i$ (hereafter referred to as the LOS velocity) is calculated using the definition $v_i = c(z_i -z_{\rm proto})/(1+z_{\rm proto})$ (\citealt{Harrison1974, Danese1980}), which is shown in Figure~\ref{fig:3ddist}.
Though the number is limited, the cumulative distribution of the LOS velocity is similar to a cumulative Gaussian distribution (see the top-right panel in Figure~\ref{fig:3ddist}).

Based on the derived LOS velocities, we derive the velocity dispersion, which is estimated by using two statistically different methods: (1) the biweight scale\footnote{As discussed in \cite{Ruel2014}, the biweight scale in \cite{Beers1990} (Eq.(9) therein) needs to be modified for being unbiased. 
Following earlier works, e.g., by \cite{Mosteller1977}, the correction increases the dispersion measurements by $\simeq 3\%~ (i.e., \sqrt{n/(n-1)}$ for $n=20)$ in our case.} and (2) the gapper estimator. 
The reason for testing two estimators is due to the limited number of available redshifts (which is $n=12$, for CO detection), in light of the discussions addressed in \cite{Beers1990} and \cite{Girardi1993} that the gapper estimation is preferable for systems with fewer ($n\simeq10$) redshift information.

The velocity dispersion is $\sigma_{\rm los, CO} = 421\pm148$ (biweight) and $365\pm 128$ (gapper) km s$^{-1}$. The errors are derived by taking the root sum of the squares of the statistical error calculated following the estimate in \cite{Ruel2014} and the error of 80 km s$^{-1}$ coming from the assumed spectral resolution.
With 12 CO-detections, individual estimates agree well with each other within the errors.
Therefore, we adopt the biweight scale for the order estimate of the protocluster mass in the following.
We note that the derived velocity dispersion of the protocluster including all redshift information is $\sigma_{\rm los} = 451$~km s$^{-1}$ (biweight) and $\sigma_{\rm los} =  498 $~km s$^{-1}$ (gapper); this is consistent with CO-only base measurements though the errors can be much larger (up to $\approx600$ km s$^{-1}$) than the CO-based estimates, caused by the coarser spectral resolution of the grism spectroscopy.

We finally derive the halo mass of the protocluster from the derived velocity dispersion, using two different calculations used in \cite{Finn2005} and \cite{Evrard2008}.
\cite{Finn2005} derived the halo mass based on the modified formulation of the virial mass at a given virial radius, which is expressed as
\begin{equation}
M_{\rm cl} = 1.2\times 10^{15} \big(\frac{\sigma_{\rm los}}{1000~{\rm km~s^{-1}}}\big)^3 \frac{1}{\sqrt{\Omega_{\Lambda} + \Omega_{0} (1+z)^3}}~h^{-1}_{100} ~M_{\odot}.
\end{equation}
where, $h_{100} = H_0/100$~km s$^{-1}$ Mpc$^{-1}$.
\cite{Evrard2008} derived the halo mass based on a model fit between the (DM) velocity dispersion and $M_{\rm cl}$ in the DM-only simulation expressed in the following formula
\begin{equation}\label{eq:dmcl}
M_{\rm cl} = 10^{15} \Big [\frac{\sigma_{\rm DM} (M,z)}{\sigma_{\rm DM, 15}}\Big]^{1/\alpha}h(z)^{-1}~M_{\odot}
\end{equation}
where $\sigma_{\rm DM, 15}$ is the normalization at mass 10$^{15}$~ $h^{-1}~M_{\odot}$, $\alpha$ is the logarithmic slope, and  $h(z) = H(z) /100$~km s$^{-1}$ Mpc$^{-1}$. 
The best-fit values are $\sigma_{\rm DM, 15} = 1082.9\pm4.0$~km s$^{-1}$ and $\alpha = 0.3361 \pm 0.0026$ from the simulation, which we adopt for our calculation.
We use the (biweight) velocity dispersion to be equivalent to $\sigma_{\rm DM}$ for this calculation.
We note that Equation~(\ref{eq:dmcl}) is not changed even when baryonic physics are included (\citealt{Munari2013}).

The range of halo masses from these two methods is $\log{(M_{\rm cl}/M_{\odot})} =$13.4-13.6 $\sim \mathcal{O}(10^{13})$, though the value should be taken as an upper limit for an unvirialized system. 
The progenitors of Virgo-like clusters ($M_{\rm halo}\sim (3-10)\times10^{14}~M_{\odot}$) or the most massive one like Coma clusters ($M_{\rm halo} >10^{15}~M_{\odot}$) at $z\sim2.5$ are expected to have a halo mass of $> 10^{13}~{\rm M}_{\odot}$ in simulations (\citealt{Chiang2013}).
Therefore, if the protocluster follows the typical evolutionary track as the simulation expects, our estimate suggests that the protocluster is likely to become a Virgo-like, intermediate-mass cluster at $z=0$ or even a more massive cluster.\\

When discussing the halo masses of protoclusters, one should bear in mind that there is a large uncertainty regarding the assumption of virialization as a whole.
This is because protoclusters are by definition, unvirialized, premature clusters at high redshift, though many studies have assumed they were (locally) virialized to estimate their halo mass (e.g.  \citealt{Shimakawa2014, Kubo2016, Miller2018}).
Measurements based on X-ray detection or Sunyaev-Zel'dovich (SZ) effect can be alternative methods to estimate the halo mass, but we do not have deep-enough exposures to test whether the hot intracluster medium due to virial shocks is already in place.
While deeper X-ray or SZ follow-up observations are required, we take a different approach in the next section to alleviate the uncertainties related to the estimated halo mass by evaluating the mass of the substructures of the protocluster that might be locally virialized.

\subsection{Halo mass of individual substructures}\label{sec:subhalo}
Protocluster 4C~23.56 is known to have two density peaks in the projected map (\citealt{Tanaka2011}; \citetalias{minju2017a}) and hence, the protocluster center is difficult to define.
Based on this distribution, we define two clumps around these two peaks, namely, the East Clump (EC) and the West Clump (WC).
The sizes of these clumps are defined in the projected map using the expected typical protocluster core radius ($\langle R_{200}\rangle$) at $z\sim2$, which is expected to evolve into a core of of $M_{\rm cl} > 10^{14}~M_{\odot}$ cluster at $z=0$ from simulations (\citealt{Chiang2017}). 
The radius is 0.8 Mpc in comoving scale or $\sim0.2$~Mpc in physical scale.

The EC includes seven HAEs (HAE~2, 4, 6, 9, 13, 18 and 19) with three (five) CO(CO+H$\alpha$) spectroscopic redshifts, while the WC includes nine HAEs (HAE~1, 7, 8, 10, 12, 16, 20, 21 and 23) with seven (eight) redshifts.
With the increased redshift information, we now confirm that two substructures are kinematically connected with only a small systemic velocity offset of $\approx 220 (40)$~km s$^{-1}$ if limited to galaxies with CO (CO+H$\alpha$) redshifts.
The small difference implies two galaxy groups in a merging phase within the protocluster.

Using the same method as addressed in the previous section, the estimated halo masses for both clumps are $\log{(M_{\rm 200}/M_{\odot})} = {13.0}$ and ${13.8}$ for the EC and the WC from $\sigma_{\rm EC, CO-only}= 324 \pm 274$~km s$^{-1}$ and $\sigma_{\rm WC, CO-only}=567 \pm 171$~km s$^{-1}$, respectively.
Therefore, it is $\sim \mathcal{O}(10^{13})$, which is consistent with the estimate in the previous section. 
The uncertainty is larger due to the smaller number of CO detections per clump.

We can derive the expected virial radius based on velocity dispersion (\citealt{Finn2005}), which is 0.2-0.4 Mpc in our case. 
Therefore, we may reasonably conclude that each clump defined by a projected radius of $\sim0.2$~physical Mpc is virialized locally.
The estimates of halo masses for individual clumps support the idea that the 4C~23.56 protocluster is composed of two virialized groups, which are likely to merge and may evolve into a present-day Virgo-like cluster.
Such merging of smaller scale groups could be more common for protoclusters, considering the merger trees of present-day clusters.\\

As a final remark for this section, we note two additional uncertainties that were not fully taken into account in the estimate of the halos mass and the potential descendant.
First, the order estimate of the protocluster is not based on the ``complete samples" in terms of galaxy populations. 
Here, we rely on the H$\alpha$-selected galaxies, which are identified in the flux-limited observations, and the follow-up observations of the CO lines are also flux limited.
We do miss less massive galaxies and quenched populations, if any, without significant star-forming activity.
They might be distributed differently over the protocluster region, which may change our interpretation of mass of the protocluster and the descendant, as well as their evolutionary picture (\citealt{Muldrew2015}).
In this regard, we are also limited by the survey area wherein the entire structure of the protocluster can be extended up to $R_e\sim5-10$ Mpc comoving from simulations (e.g., \citealt{Chiang2013}).
Second, we rely on simulated results in which scatter is large, as shown in \citet[ see Figure.~2 therein]{Chiang2013} though this is our best conjecture on the descendant of the protocluster in the local universe.
We may be able to solve the problem observationally be collecting a larger number of (proto)clusters and assessing them statistically in terms of their number density as a function of redshift. 

\section{Galaxy kinematics in context}\label{sec:kinematics}
\subsection{Mergers in the 4C~23.56 protocluster} \label{sec:mergers}
In Section~\ref{sec:subcomponent}, we found that HAE~9 and HAE~16 have subcomponents that are likely associated with them. Here, we present further discussions on whether HAE~9 and HAE~16 are experiencing gas-rich mergers, based on ancillary data sets.

For HAE~9, we utilize the misalignment diagnostics (\citealt{Franx1991}). 
We measure the difference between the geometrical position angle (PA$_{\rm geo}$) and the kinematical position angle (PA$_{\rm kin}$).
The kinematical position angle is defined by the line that connects the peak positions of the blue (main) and red (sub) components in HAE~9.
We used the 1.1 mm continuum as a probe of PA$_{\rm geo}$, where the emission is extended and marginally resolved at $0''.78\times0''.68$ resolution.
We used CASA task $\mathtt{uvmodelfit}$ for measuring PA$_{\rm geo}$, giving PA$_{\rm geo}=-60^{\circ} \pm 19^{\circ}$ (shown as a white straight line in Figure~\ref{fig:merger_hae9}).
The misalignment between the morphological and the kinematical PAs for HAE~9 is $\approx 30^{\circ}$.
As a reference, the misalignment angle between the morphological and kinematic PAs for disk-like star-forming galaxies at high redshift is typically less than $30^{\circ}$ with a median of $12^{\circ}$ (e.g.  \citealt{Wisnioski2015}).
The misalignment could have originated from external forces either from a merger, gas stripping, or galaxy harassment in the protocluster.

There is supporting evidence that the misalignment originated from a gas-rich merger.
The line width of the CO~(3--2) emission, observed at $0''.91\times0''.66$, exceeds 700 km s$^{-1}$, which is barely seen in a typical isolated galaxy.
Further, the 1.1 mm flux was one of the highest among the detections \citepalias{minju2017a}, close to the flux of dusty star-forming galaxies (DSFGs) or submillimeter-bright galaxies (SMGs), some of which are found to be mergers.
The peak position for the subcomponent is offset by $0''.34$ from the peak of the main component, which is larger than the minor axis of the beam, implying that these are perhaps distinct galaxies.
The $K_{\rm p}$-band AO image for this galaxy, using the IRCS/Subaru, shows two separable components in the smoothed map (\citetalias{minju2017a}), further supporting a merger origin for the CO emissions in HAE~9.

What about HAE~16? We searched for the multiwavelength counterparts of the companion, including the CO~(3--2) and the near infrared data sets in hand.
Unfortunately, we could not confirm the existence of a counterpart at other wavelengths.
Nevertheless, this does not reject the minor gas-rich merger scenario because observations are not deep enough to detect galaxy mergers with the stellar mass ratio less than $\sim1:10$. 
The 3~$\sigma$ limiting magnitudes of the $Ks$ and $J$ bands with Subaru/MOIRCS are 24.47 (AB magnitude) and 24.79 (AB magnitude), respectively, and the depth may have detected galaxies with stellar mass as low as $\log({M_{\star}/M_{\odot}})\sim 9.5$ in principle, though the estimate can vary with assumptions of the IMF, metallicity, and star-formation history.
If there is a galaxy lower than this stellar limit, it would have been missed with the current Subaru observations.
Also, previous CO~(3--2) observations lacked the point-source sensitivity of the CO~(4--3) observations.
We need deeper observations to confirm the existence of a gas-rich minor merging counterpart.
However, the gas-rich nature ($f_{\rm gas}\approx0.8$, \citetalias{minju2017a}) of HAE~16 may be reasonably explained by gas-rich accreting materials considering the existence of metal-enriched, gas-rich component accreting from (or merging with) the outer filaments of the large-scale structure (\citealt{Emonts2016, Ginolfi2017}).

\begin{figure*}
\centering
\includegraphics[width=0.9\textwidth, bb = 0 0 1800 800]{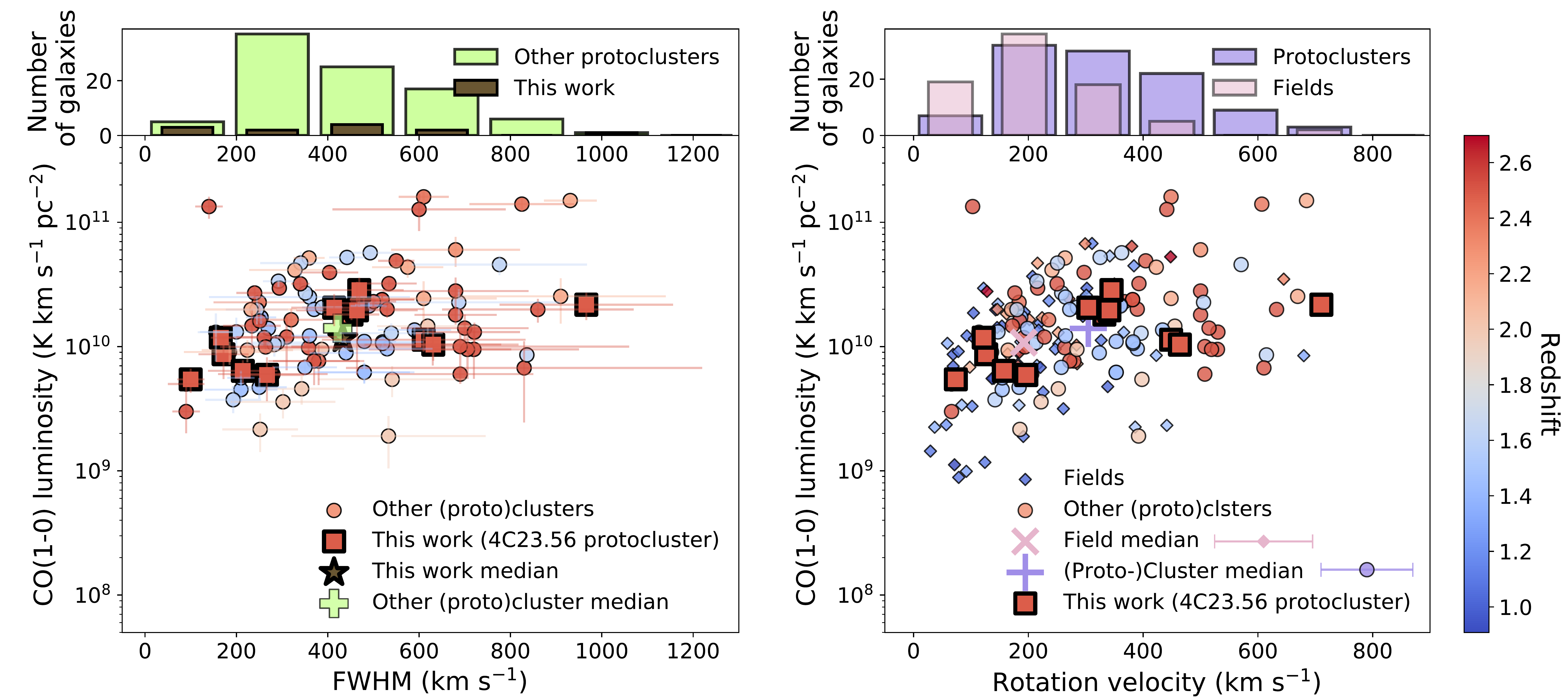}
\caption{Left: the distribution of the CO~(1--0) luminosity and CO line widths obtained from various $J$ transition (from $J=1$ up to $J=4$) observations. We converted to CO~(1--0) luminosity by assuming the line luminosity ratio of $R1J (=L'_{\rm CO1-0}/L'_{\rm COJ-(J-1)}) = 1.2, 1.8, {\rm and}~2.4$ for $J= 2,3, {\rm and}~4$, respectively. We retrieved the values from \cite{Aravena2012, Casasola2013, Ivison2013, Tadaki2014b, Tadaki2019a, Chapman2015, Hayashi2017, Rudnick2017, Dannerbauer2017, Noble2017, Noble2019, Wang2018, Coogan2018} and \citet[circles]{Gomez-Guijarro2019} in addition to our CO measurements (squares). Different colors of these symbols correspond to different redshifts as shown in the color bar on the right side. Right: the distribution of CO~(1--0) luminosity and rotation velocity in the comparison with field galaxies. Field galaxies are obtained from \citet{Tacconi2013, Daddi2015, Bourne2019, Aravena2019}. The rotation velocity is converted from the FWHM by taking the isotropic virial estimate of the circular velocity (FWHM$\times \sqrt(3/8\ln{2})$ to match with galaxies in \citet{Tacconi2013}, which provides the largest number for field population ($n=52$). Again, different colors of the symbols show the redshifts. The median values of the CO luminosity and rotation velocities for the field and protocluster galaxies are shown by ``X" and ``+" symbols, respectively, and we show the typical errors of the rotation velocity on the right of the corresponding legend. ln each panel, we show a histogram of the FWHM values or the rotation velocity. \label{fig:fwhm}}
\end{figure*}

\subsection{CO line widths and $L'_{\rm CO}$}\label{sec:comparison}
For the discussion of the remaining CO-detected galaxies, we compare the CO line widths (FWHM) and CO luminosities ($L'_{\rm CO}$) with other CO studies both in (proto)clusters and general fields. 
The line width is sensitive to both dynamical mass and any inclination effects, while $L'_{\rm CO}$ is sensitive to the total mass of molecular gas.
We discuss the characteristic properties of 4C~23.56 protocluster galaxies based on this.

For protocluster galaxies, we obtained the measurements from \cite{Aravena2012, Casasola2013, Ivison2013, Tadaki2014b, Chapman2015, Rudnick2017, Dannerbauer2017, Hayashi2017, Noble2017, Noble2019, Wang2018, Coogan2018, Tadaki2019a, Gomez-Guijarro2019}. 
All of these studies targeted galaxies associated with protoclusters at $1.5\lesssim z\lesssim 2.5$ and detected low-/mid-$J$ CO lines (up to $J=4$).
If the SFR and stellar mass measurements are available, we also checked these values. 
Galaxies from the (proto)cluster studies are scattered around the main sequence, while some galaxies ($9/84$, 84 corresponds to the number of galaxies with $M_{\star}$ and SFR in the literature) have exceptionally high SFR $\gtrsim 500~M_{\odot}$ yr$^{-1}$ (or the corresponding $L_{\rm IR}$) that largely deviate from the main sequence.

There are several measurements for the same galaxies in different angular resolutions and/or different CO lines in the protocluster studies.
We used the CO~(1--0) measurements in \citet{Tadaki2014b} and  \citet[ $n=2$ and $1$, respectively]{Dannerbauer2017} instead of taking the values from the CO~(3--2) measurements in \citet{Tadaki2019a}.
For \citet[$n=4$]{Noble2017, Noble2019}, we used the updated values in \citet{Noble2019} for the CO~(2--1) line detection.
For overlapping sources between \citet{Wang2018} and \citet[$n=10$]{Gomez-Guijarro2019}, we used the CO~(1--0) measurements from \citet{Wang2018}. 
We note that the line widths in different transitions are comparable with each other for these works.
For \citet{Coogan2018}, we used the CO~(4--3) measurements instead of those for CO~(1--0), due to the low S/N in CO~(1--0).
Adding our CO measurements of CO~(4--3) and CO~(3--2) (i.e., 11 from CO~(4--3) and one from CO~(3--2)), a total of 103 galaxies are considered (proto)cluster galaxies.

We compiled the field data points from \citet{Tacconi2013, Daddi2015, Bourne2019} and \citet{Aravena2019}.
For field galaxies, we filtered out one galaxy at $z>3$, yielding the field control sample of $n=80$ at $1\lesssim z \lesssim3$.
We note that these field galaxies are also located around the main sequence, except for one with an SFR of $630~M_{\odot}$ yr$^{-1}$.

If we only focus on the CO line widths, the median value for the 4C~23.56 protocluster galaxies is 433 km s$^{-1}$, compared to 422 km s$^{-1}$ for other protocluster galaxies, which is indicated in the left panel of Figure~\ref{fig:fwhm}.
Considering the typical errors for fitting and assumed spectral resolution ($\sim100$ km s$^{-1}$), we conclude that CO line widths are consistent with each other.
The result does not change even if we take all CO measurements for overlapping sources.

For a fair comparison with field galaxies, we converted all CO line widths(=FWHM) to the rotation velocity by taking the isotropic virial estimate of the circular velocity (FWHM$\times \sqrt{3/(8\ln{2}})$ to match with galaxies in the PHIBSS survey (\citealt{Tacconi2013}), which provides the largest number of field galaxies near the main sequence ($n=52$).
The authors reported the characteristic rotation velocity ($v_{\rm rot}$) either from the line FWHM for unresolved sources or the inclination-angle-corrected velocity gradient for resolved sources.
Because the line FWHMs, the velocity gradients, and the inclination angles are not listed in their tables,
we apply the prescription to unresolved cases for 103 protocluster galaxies and for the remaining field galaxies ($n=28$). 
All data points are shown in the right panel of Figure~\ref{fig:fwhm}.

There is a hint of higher rotation velocity (from the CO line widths) for protocluster galaxies.
We find a higher median value for the protocluster ($v_{\rm rot, pc}$= 304 km s$^{-1}$) compared to the field galaxies ($v_{\rm rot, fd}$= 192 km s$^{-1}$), which is shown in Figure~\ref{fig:fwhm}. 
There is a clear tail toward higher rotation velocity in the histogram for protocluster galaxies.
Such a difference becomes even larger if we exclude galaxies from \cite{Tacconi2013}.
Considering random inclination angles that affect the observed line widths for both field galaxies and protocluster galaxies, the difference between protocluster and field galaxies is likely the intrinsic feature.

We also converted the CO line flux into the CO~(1--0) luminosity as a probe of molecular gas mass in protoclusters.
We assumed the line luminosity ratio of $R1J (=L'_{\rm CO1-0}/L'_{\rm COJ-(J-1)}) = 1.2, 1.8,$ and $2.4$ for $J= 2,3, {\rm and}~4$, respectively, which are the values often adopted for typical star-forming galaxies (e.g., \citealt{Dannerbauer2009, Aravena2010, Bolatto2015, Daddi2015}).
While the CO emissions may be highly excited compared to field galaxies for some of the galaxies  (e.g., \citealt{Coogan2018}),
we take the nominal ratios as we do not have secure constraints for many of the protocluster galaxies. This is why we tried to use the CO line width and the flux from CO~(1--0) measurements, when available.
We also note that the depth of field and protocluster studies varies, which makes a fair comparison difficult. 
However, most observations except for those targeting extreme starbursts have similar depth in terms of the $L'_{\rm CO~(1-0)}$ limit, i.e., one to a few $\times10^{9}$ (K km s$^{-1}$ pc$^{-2}$). 
Therefore, the general discussions below would not be largely affected by the different sensitivities of the  different studies that we consider.

In terms of the CO~(1--0) luminosity, we find that there is 0.11 dex difference in terms of the median value between the 4C~23.56 protocluster ($\log{(L'_{\rm CO10}}$/$({\rm K\,km\,s^{-1}\,pc^{-2}))} = 10.04$) and other protocluster galaxies ($\log{(L'_{\rm CO10}/({\rm K\,km\,s^{-1}\,pc^{-2}}))}=10.15$), which is negligible considering the typical errors.
The consistency does not change even if we take all available CO measurements.
Therefore, we conclude that all (proto)cluster galaxies that have been considered have consistent line widths and the CO~(1--0) luminosities on average.

Similarly, the difference between field galaxies and all protocluster galaxies in terms of $L'_{\rm CO~(1-0)}$ is also negligible ($\log{(L'_{\rm CO10}}/$(K km s$^{-1}$ pc$^{-2})) = 10.13$ versus $\log{(L'_{\rm CO10}/({\rm K\,km\,s^{-1}\,pc^{-2}))}} = 10.01$). 
This suggests that we do not see a higher gas fraction per galaxy on average for protocluster galaxies that are mostly on the main sequence, confirming our previous argument in \citetalias{minju2017a}.

There are two possible ways to explain the difference in the line width.
First, protocluster galaxies are kinematically different, perhaps owing to higher merger rates, which are unresolved with the large beam size, broadening the line width.
Related to this aspect, it is worth noting that most of these measurements are based on observations at spatial resolutions of $\approx1''-6''$.
Therefore, most sources are unresolved.
Second, protocluster galaxies have intrinsically large line widths, suggestive of larger dynamical masses or different mass distributions.
The current sensitivity is not sufficient to conclude whether one dominates over the other, or both contribute equally.

Regarding the possibility of unresolved mergers, several protocluster studies have reported enhanced merger rates based on rest-frame optical images compared to fields (e.g., \citealt{Hine2016, Coogan2018, Watson2019}).
Here, the kinematical property of a larger line width may consistently indicate that many protocluster galaxies are undergoing (gas-rich) mergers that are not spatially resolved.

For the 4C~23.56 protocluster, indeed, HAE~9 has a broader line width in the CO~(3--2) emission, which was not resolved with the large beam ($\sim0''.8$).
The CO~(4--3) observations at higher angular resolution suggest (Figure~\ref{fig:merger_hae9}) that this galaxy may be a gas-rich merger (Section~\ref{sec:subcomponent} and \ref{sec:mergers}).
Other potential merger candidates with large line widths (FWHM$>400$ km s$^{-1}$; Table~\ref{tab:detection}) from the CO~(4--3) detections are HAE~1, HAE~3, HAE~7, HAE~10, and HAE~23.
We discussed in \citetalias{minju2017a} that HAE~3 may be a merger from its resolved stellar map.
For HAE~1 and HAE~7, there are hints of elongated emissions in CO~(4--3) as indicated in Figure~\ref{fig:gal2},  which needs future deeper observations to confirm whether they are `very' close pairs or outflowing components. 
The current sensitivity limits such confirmation.

For the second possibility of an intrinsically large line width, it is important to constrain the size of the galaxies first.
We fitted the CO intensity with an elliptical Gaussian using the CASA task imfit for the 4C~23.56 protocluster galaxies.
We could only constrain the sizes of the galaxies for HAE~8, 9 (main galaxy), 16 (main galaxy) from the fitting, yielding semi-major beam-deconvolved sizes of $0''.21-0''.31$ (with errors $0''.08-0''.12$), meaning that the galaxies are marginally resolved with the beam ($0''.52\times0''.32$).
This corresponds to $\approx1.7-2.6$ kpc.
The S/Ns for other galaxies are insufficient to make useful measurements.

 \citet{Noble2019} estimated the CO~(2--1) sizes with a beam size of $\sim0''.4$, yielding $R_{\rm 1/2,CO} \sim 2-7$ kpc.
They argued for smaller sizes in CO emission for cluster galaxies at $z=1.6$ compared to field galaxies, though the statistical difference was small ($\sim0.9\sigma$).
For other cluster galaxies at slightly higher redshift $z=1.99$, \citet{Coogan2018} also measured the size, giving the CO~(4--3) size constraints for two galaxies with $FWHM/2\approx 0''.2-0''.3$, comparable to our measurements, and the upper limit for three galaxies ($<0''.16$).

The number of galaxies with CO size constraints is also limited for field galaxies at $z\gtrsim2$.
Even from the PHIBSS, we could only obtain one value from the list, which is 3.8 kpc, estimated from the CO~(3--2) line. 
In \citet{Aravena2019}, they measured the CO~(3--2) emitting sizes from two galaxies at $z\sim2.5$ using (see their Appendix C), which is $\approx 4~$kpc from the Gaussian fit.
The CO sizes of HAE~8, 9, and 16, are smaller than those of the field galaxies at similar redshift.

If we take the size and line width differences as a generic feature of protocluster galaxies with caveats of inhomogeneous selection and small sample sizes, 
this suggests that the dynamical mass ($\propto rv_{\rm rot}^2$) traced by CO lines is comparable to or slightly higher (by $\sim20\%$) than the field galaxies, and the mass is more centrally concentrated for protocluster galaxies.
We note that the above field and cluster galaxies have stellar masses comparable to HAE~8, 9, and 16 and thus the size difference may not be due to the difference in the stellar masses.
Different CO transitions can have different emitting sizes, but we compare close CO transitions at similar redshifts, and the low-$J$ and mid-$J$ CO emissions are known to have similar distributions, at least for typical main-sequence galaxies (e.g., \citealt{Bolatto2015}).

It may be worth noting that two main galaxies of HAE~9 and HAE~16 are likely to be undergoing gas-rich mergers and have smaller CO sizes ($\sim2\times$) compared to the field.
Considering this, two scenarios that explain the large line width for protocluster galaxies may not be independent but intertwined mechanisms. In other words, mergers may also contribute to the intrinsically large line width with smaller sizes by redistributing the gas.

In Section~\ref{sec:halos}, we found that the 4C~23.56 protocluster is kinematically composed of two merging galaxy groups that are loosely connected.
In such environment where hierarchical structure formation is underway, galaxy mergers should also be prevalent, where the velocity dispersion between galaxies is smaller compared to local virialized clusters.
HAE~16 and HAE~9 are indeed located in two dense substructures of the protocluster (see Figure~\ref{fig:3ddist} and Section~\ref{sec:halos}) with moderate galaxy--galaxy dispersions (i.e., $324$ and $567$ km s$^{-1}$), which increase the probability of mergers.

\begin{figure}
\centering
\includegraphics[width=0.48\textwidth, bb=0 0 1000 900]{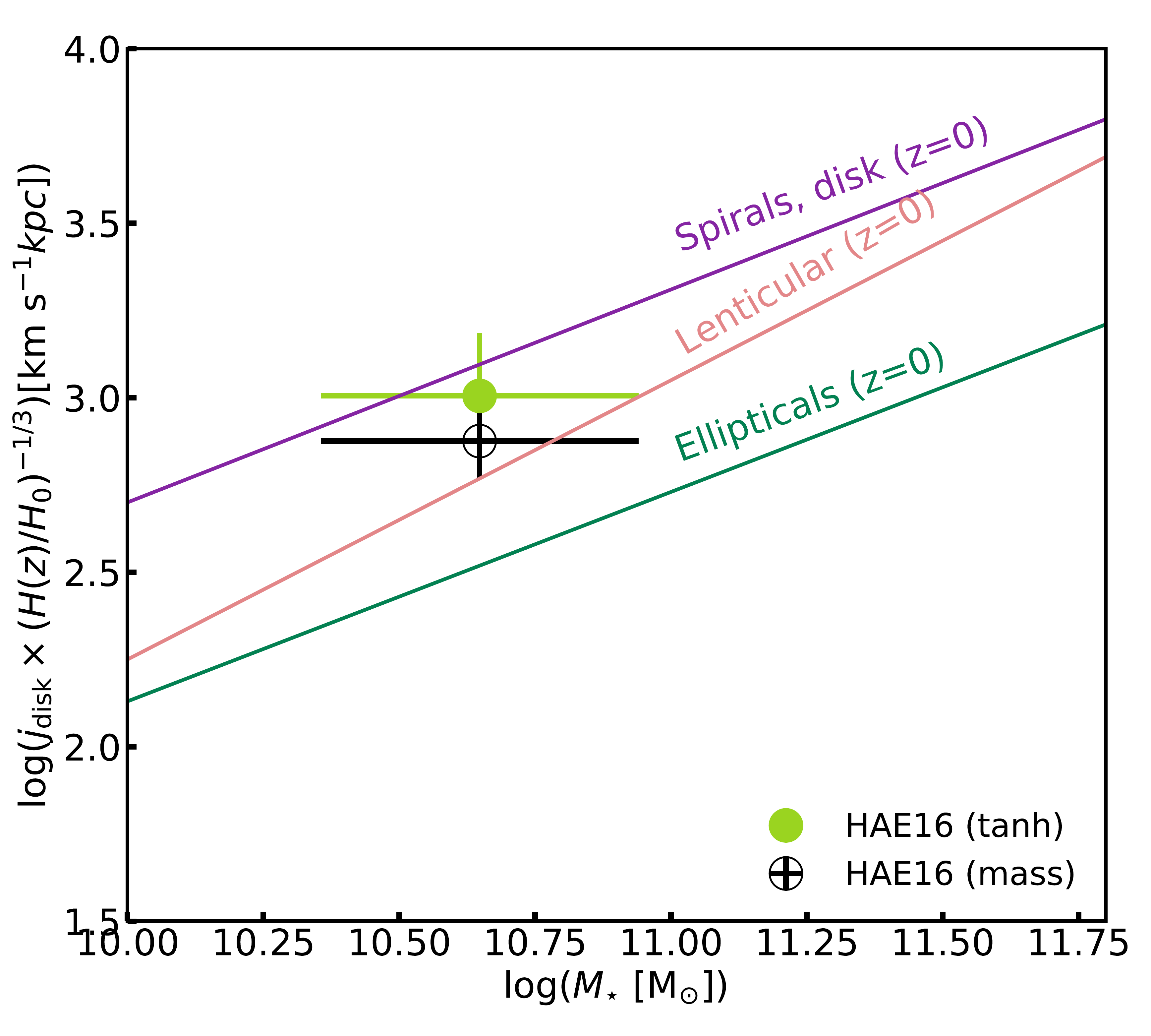}
\caption{The specific angular momentum of HAE~16. For comparison, we plotted the empirical fit presented in \cite{Romanowsky2012}. We assume that the specific angular momentum of the disk is comparable to that of stars. \label{fig:AM}}
\end{figure}

If mergers are frequent, then the next question would be the role of these mergers in the formation of ETGs. 
In Figure~\ref{fig:AM}, we plot the local empirical relations between the specific angular momentum and the stellar mass for the three morphological types of spiral, S0, and elliptical galaxies from \cite{Romanowsky2012}. 
\cite{Fall1980} found that elliptical and spiral galaxies form parallel sequences, with the former having a factor of $\approx 6$ smaller specific stellar AM ($j_{\star}$) than the latter.
\cite{Romanowsky2012} updated the work from \cite{Fall1980} and showed that galaxy mergers can naturally explain the positions of elliptical galaxies in the $j_{\rm stars}-M_{\rm stars}$ plane 
and that disks and bulges follow fundamentally different $j_{\rm stars}-M_{\rm stars}$ relations. 

In light of this, we derive a crude estimate of the specific angular momentum ($j_{\rm disk} = 2\times v_{\rm circ}\times R_{\rm disk} = 1.19\times v_{\rm circ}\times r_e$, where $v_{\rm circ}$ is circular velocity, e.g., \citealt{Romanowsky2012, Obreschkow2014, Burkert2016}) based on the size and ine widths for field and protocluster galaxies. 
It gives a value comparable to or slightly smaller (by $\approx 30\%$) than that for protocluster galaxies than the field star-forming galaxies in general.

For HAE~16, we could measure the specific angular momentum of the disk from kinematical modeling with GALPAK$^{\rm 3D}$ (\citealt{Bouche2015}; see Appendix~\ref{app:galpak} for the detailed kinematic modeling of HAE~16 and discussions on the derived values). 
While the galaxy may have a gas-rich companion, it shows a smooth velocity gradient that allows us to model the galaxy. 
The estimated specific angular momentum after correcting for the redshift dependence of the Hubble constant in log scale are in a range between 2.8 and 3.0 (Figure~\ref{fig:AM}).
For the circular velocity, we took into account the pressure component from turbulent motion, which makes it larger than the rotation velocity $v_{\rm rot}$ alone, $v_{\rm circ}^2 = v_{\rm rot}^2 + 2 \sigma_0^2 r_e/R_{\rm disk}$ (e.g., \citealt{Ubler2017}). 
For our case, this increases the specific angular momentum value by 6-16\%, depending on the assumed velocity profiles for fitting.
We consider that the stellar specific angular momentum of the local universe is comparable to the (cold) gas specific angular momentum (e.g., \citealt{Tadaki2017b}), because high-$z$ galaxies are gas rich ($f_{\rm gas} \sim 0.5)$.

Our findings for HAE~16 suggest that this galaxy was, at the very least, born in a similar way to the local spirals. Even if the galaxy might have experienced mergers, the specific angular momentum has not decreased. The galaxy must lose a significant amount of angular momentum via ``dissipational" processes to become local elliptical galaxies that are abundant in cluster regions.

\cite{Lagos2018} recently presented a quantified assessment for the impact of mergers in galaxy evolution, including dry, wet, minor and major mergers using the EAGLE simulation.
They found that regardless of minor or major mergers, ``wet" mergers tend to increase the specific angular momentum, thus spinning up the galaxy rotation, while ``dry" mergers tend to decrease the specific angular momentum.
Even if additional dissipative processes from the environment is in acting, the angular momentum might not change but might remain similar values to field galaxies, perhaps with compact rotating star-forming cores at high redshift (e.g., \citealt{Barro2017b, Tadaki2017b, Talia2018}).
This picture is also consistent with Illustris simulation in \citet[see also \citealt{Teklu2015}]{Genel2015} that ETGs -- including slow rotators, which are round and have lower angular momentum at $z = 0$ -- reside at high redshift on the late-type (spiral) relation and need active galactic nucleus (AGN) feedback or gas-poor mergers to lose the angular momentum.

The next step to confirm mergers and to quantify the exact role of gas-rich mergers is to have higher angular resolution observations with good sensitivity.
As the number of CO-detected galaxies in fields and protoclusters has now reached $\sim100$ at high $z (z>1)$, 
we need to verify the above scenario with a larger number of samples.

\section{Summary}\label{sec:conclusions}
In this paper, we investigated kinematical properties of star-forming galaxies on the main sequence associated with the protocluster 4C~23.56 at $z=2.5$ with the CO~(4--3) line at $\sim 0''.4$ resolution using ALMA.
We summarize our findings and discussions as follows:
\begin{enumerate}
\item We detected CO~(4--3) line emissions from 11 out of the 16 HAEs that were targeted. These include six HAEs that were previously detected in CO~(3--2) line emissions within the covering FoVs. The number of CO-detected galaxies increased by a factor of 2.
The simultaneous detections in the CO~(3--2) and CO~(4--3) (HAE~3, 4, 8, 9, 10, and 16) lines were consistent in terms of the line widths and the redshifts, when spatial and spectral resolutions are matched, confirming both line detections.

\item We estimated the halo mass of the protocluster based on the increased redshift information from the CO lines.
The estimated halo mass under the assumption of local virialization is $\log{(M_{\rm halo}/M_{\odot})}=13.4-13.6$.
Comparing with the cosmological simulations in \citealt{Chiang2013}, the protocluster is likely a progenitor of a present-day Virgo-like cluster ($>10^{14}~M_{\odot}$), though there are limitations for the estimate. We also confirmed two galaxy groups within the protocluster that are likely merged into one system.

\item Two gas-rich galaxies ($f_{\rm gas}>0.7)$, HAE~9 and HAE~16, have potential subcomponents or potential merging counterparts. These components are resolved spatially and spectroscopically in the CO~(4--3) line. 
A less luminous subcomponent is found 2.5 kpc westward of HAE~9 with a velocity offset of $\sim500$ km s$^{-1}$, which was not resolved in our previous CO~(3--2) observations. 
A potential gas-rich minor merging counterpart is found at 10 kpc apart from HAE~16, with a velocity offset of $\sim 100$ km s$^{-1}$.

\item With increased CO detections ($n=12$) for our protocluster galaxies, we compared their CO line width and the CO~(1--0) luminosity with those of other protocluster galaxies ($n=91$) and field populations ($n=80$). We found that the line widths and the CO luminosities are comparable between the 4C~23.56 protocluster and other protoclusters (i.e., FWHM$_{\rm median} = 433$ (4C~23.56) and $422$ (other protoclusters) km s$^{-1}$ and $\log{(L'_{\rm CO10}}/$ (K km s$^{-1}$ pc$^{-2}$)), median $= 10.04$ (4C~23.56) and $10.15$ (other protoclusters). On the other hand, the CO line widths for protocluster galaxies are broader on average compared to field galaxies, i.e., $v_{\rm rot, median} = 304$ (protocluster galaxies) and 192 (field galaxies) km s$^{-1}$, while the median CO luminosities are comparable in different environments.
\item Based on our resolved views for our protoclusters, which have gas-rich mergers (HAE~9 and HAE~16) and smaller sizes (HAE~8, 9 and ~16 of $\approx2$ kpc), we conclude that broader line widths (or the rotation velocity) may be attributed to effects of both unresolved mergers and compact gas distribution for (at least) the 4C23.56 protocluster.
\end{enumerate}

Based on the comparison of recent cosmological simulations to observed results, gas-rich mergers may play a role in the retention of the specific angular momentum, similar to the field population, during cluster assembly. 
We tested this with one galaxy (HAE~16), which is likely undergoing a gas-rich merger, and found that its specific angular momentum comparable to local spirals. This suggests that specific angular momentum needs to be decreased in later times if a galaxy evolves into a local elliptical galaxy (or a slow rotator), which is abundant in clusters.
Additional dissipational processes including gas-poor mergers and AGN activities may be necessary and might play such a role.
Future deeper observations equipped with higher angular resolution are highly desirable for constraining the physical processes in kinematical transformation in high-redshift overdense regions and the role of environments, if any.

\acknowledgments

We are immensely grateful to the anonymous referee for constructive comments.
This work was supported by NAOJ ALMA Scientific Research Grant Number 2018-09B.
This research made use of astrodendro, a Python package to compute dendrograms of astronomical data (http://www.dendrograms.org/). We thank Sedona Price and Sirio Belli for fruitful discussions on the CO line widths, sizes, and dynamical masses.
This paper makes use of the following ALMA data: ADS/JAO.ALMA \#2015.1.00152.S. ALMA is a partnership of ESO (representing its member states), NSF (USA) and NINS (Japan), together with NRC (Canada) and NSC and ASIAA (Taiwan) and KASI (Republic of Korea), in cooperation with the Republic of Chile. The Joint ALMA Observatory is operated by ESO, AUI/NRAO, and NAOJ.
Y.T. is supported by JSPS/MEXT KAKENHI (No.\ 17H06130).
Data analysis was in part carried out on the open-use data analysis computer system at the Astronomy Data Center, ADC, of the National Astronomical Observatory of Japan.
\vspace{5mm}
\facilities{ALMA, Subaru}.


\software{
CASA \citep{McMullin2007} , 
Astropy \citep{astropy}}


\appendix

\section{CO~(4--3)-detected galaxies without the CO~(3--2) line}\label{app:newdetection}
We detect five additional galaxies in the CO~(4--3) line that were not detected in our previous CO~(3--2) observations.
They suggest higher gas excitation in CO SLEDs than those detected simultaneously in CO~(3--2) and CO~(4--3) lines, but again, we need other (lower and higher $J$) CO transitional lines and deeper observations of CO~(3--2) for confirmation. 
Instead of discussing the CO SLEDs, we present a brief summary of these newly detected galaxies, based on the available ancillary data sets and literature.
\begin{description}
\item[{\bf HAE~1}]  The CO~(4--3) emission detected from HAE~1 appears to be associated with either a radio jet or a halo component of the radio galaxy.
In Figure~\ref{fig:gal1}, we plot the position of the radio core detected in JVLA observations in the $S$ and $C$ bands (observing frequency of 3~GHz and 6~GHz, respectively; M. Lee et al. 2019, in preparation) and the direction of the bipolar radio jet together with the CO~(4--3) image.
The positional relation of the radio core and the jet relative to the CO~(4--3) emission suggests that the CO~(4--3) line emission might be part of an outflow component ejected from the radio core. 
The CO~(4--3) line is redshifted ($\approx 800$~km~s$^{-1}$) than the reference redshift estimated from H$\alpha$ emission from grism spectroscopy using Subaru/MOIRCS (\citealt{Tanaka2011}) 
and the emission lines of H$\alpha$, H$\beta$, [O\textsc{III}]$\lambda$5007, [N\textsc{II}]$\lambda$6583 from VLT/SINFONI (\citealt{Nesvadba2017}).
 \cite{Nesvadba2017} reported a broad [O\textsc{III}]$\lambda5007$ emission, 
 where the origin of the emission is attributed to the AGN, with FWHM $\simeq630$~km s$^{-1}$ on the global scale and some very disturbed regions with [O\textsc{III}] FWHM~$\approx1000$~km s$^{-1}$ in the resolved map.
The [O\textsc{III}] emission is extended over $\approx 50$ kpc ($\sim 6''$) in spatial scale.
The position of the CO~(4--3) emitting region is consistent with the redshifted [O\textsc{III}] with respect to the systemic velocity, further supporting a potential association with a potential (fast) outflow. 
Considering the energy ejected by the AGN activity, the spatial distribution of AGN-driven [O\textsc{III}] emission, and its large line width, the CO-emitting region might have been affected by the strong AGN activity, which perhaps populates higher-$J$ CO lines.
The CO~(4--3) emission without CO~(3--2) detection might imply an AGN feedback that plays a role in the surrounding ISM, which needs future confirmation.
\item[{\bf HAE~2}] The galaxy is detected in CO~(4--3) with the second lowest S/N (peak intensity $S/N\simeq 5$). 
This galaxy might be a lensed galaxy associated with the protocluster, as suggested from our $K_{\rm s}$-band AO imaging using Subaru/IRCS and our spectroscopic follow-up (I. Tanaka et al. 2019, in preparation).
Here, we only use the redshift information in the remaining discussions.
\item[{\bf HAE~7}] It is located well below the main sequence defined at $z=2.5$; the estimated stellar mass is already $\log (M_{\star}/M_{\odot})\simeq 11.47$ with sSFR = 0.21 Gyr$^{-1}$ (\citetalias{minju2017a}; I. Tanaka et al. in prep.).
The CO~(4--3) detection might imply the existence of dense molecular gas within the galaxy, but the CO~(3--2) line and the 1.1 mm continuum emissions were not detected (\citetalias{minju2017a}).
The $3~\sigma$ upper limit of gas mass set by the CO~(3--2) nondetection is log ($M_{\rm gas} [M_{\odot}])<10.93$), thus $f_{\rm gas}<0.22$ from CO~(3--2).
HAE~7 was located at the edge of the FoV for the 1.1 mm continuum, thus the sensitivity was not sufficient to securely detect the dust emission.
The uncertainties in constraining the total gas mass from CO~(4--3) may be large, due to the unknown CO excitation (i.e., detected only in CO~(4--3)), but the CO~(4--3) detection may indicate the existence of gas even at low sSFR.
With regard to this, there are several reports on the detection of cold molecular gas with low sSFR galaxies, such as in ETGs (e.g.  \citealt{Crocker2011}) or post-starburst galaxies (e.g.  \citealt{French2015}), suggesting morphological quenching (\citealt{Martig2009}).
We need further follow-up observations to reveal the nature of this particular galaxy, which may soon be quenched.

\item[{\bf HAE~12}] The CO~(4--3) line emission is offset from the peak position of H$\alpha$ by $\simeq 0^{\prime\prime}.5$.
The significance of the detection is the lowest, followed by HAE~2, and is observed near the edge of the FoV.
The galaxy is located in the vicinity of the radio galaxy (HAE~1; $\Delta \simeq 3''$) and within the extended \textsc{[OIII]}-emitting region as explained above for HAE~1. 
The CO~(4--3) emission from HAE~12 might be connected with the AGN activity of the radio galaxy 4C~23.56.

\item[{\bf HAE~23}] We detect CO~(4--3) from HAE~23 near the edge of the spectral window coverage.
The stellar mass of this galaxy is in the lowest mass range bin among 25 HAEs, i.e., $\log{M_{\star}/M_{\odot}} < 10$, where the uncertainty of the stellar mass is the largest compared to the higher mass bins.
Even if the uncertainty is taken into account, the galaxy is located well above the main sequence at the low-mass end with sSFR of 8.8 Gyr$^{-1}$ (I. Tanaka et al. 2019, in preparation).
HAE~23 is spatially located close to HAE~16 ($\Delta \approx 3''$) and the systemic velocity is different by $\sim 700$~km s$^{-1}$. 
The galaxy might be another infalling gas-rich, low-mass galaxy that may accrete onto (or be merged with) HAE~16.
\end{description}

\section{GalPak$^{3D}$ modeling and caveats}\label{app:galpak}
We present a kinematical modeling of the CO~(4--3) line for HAE~16, which is marginally resolved with the current beam; the Gaussian fitted size (=semi-major axis) is $0''.22\pm0''.08$ ($\approx1.8\pm0.7$ kpc at $z=2.5$).
HAE~16 exhibits a disk-like, smooth velocity gradient with a minor merging companion $\approx$10 kpc apart.

We note that HAE~8 was excluded for modeling even though it was detected at the highest significance in the CO~(4--3) line and exhibits a smooth velocity gradient, marginally.
We tested GalPaK$^{3D}$ modeling for this galaxy, but the kinematic parameters, including rotation velocity and inclination, did not converge.
The narrow-line width (FWHM$\simeq~110$~km s$^{-1}$), suggestive of a face-on view for the galaxy, was an obstacle in constraining the model parameters with the current spatial resolution. 
We need higher angular resolution observation for this galaxy to constrain the kinematical parameters.

We used GalPaK$^{3D}$ (\citealt{Bouche2015}, version 1.8.8), which implements the Bayesian-based Markov Chain Monte Carlo (MCMC) calculation to estimate parameters and corrects the beam-smearing effect.
We examined two types of velocity profiles provided in the package, a hyperbolic tangent (tanh) profile (e.g.  \citealt{Andersen2013}) and a ``mass" profile for modeling the galaxy.
The former reaches an asymptotic maximum velocity toward the outer disk, while the latter falls off after reaching a peak.
We used both profiles to test two characteristic velocity profiles that have been used to model high-$z$ star-forming galaxies. They are (1) an arctan profile for the asymptotic flat velocity curve (e.g., \citealt{Puech2008, Jones2010, Swinbank2012, Contini2016, Mason2017, Turner2017a}) and (2) an exponential disk profile, or Freeman disk (e.g., \citealt{Freeman1970, Binney2008}), for the falloff in outer regions (e.g., \citealt{ForsterSchreiber2006, Gnerucci2011a, Wisnioski2015, Genzel2017, Lang2017, Ubler2018}); see also \citealt{Epinat2010} for discussions on several rotation curve models used for kinematical modeling). 

The tanh profile is similar to the arctan profile and is expressed as
\begin{equation}
V(r) = V_{\rm max} \tanh(r/r_{\rm t})\,\,\,\; {\rm ``tanh"}.
\end{equation}
which is adopted to explain the flat rotation curve in the outer disk, which is often observed in local disk galaxies.

On the other hand, the ``mass" profile is described as
\begin{equation}
V(r) \propto \sqrt{\frac{I(<r)}{r}} \,\,\,\; {\rm ``mass"}
\end{equation}
and normalized by $V_{\rm max}$ in the calculation.
It assumes a constant mass-to-light ratio ($M/L$) and does not parameterize the turnover radius.
The use of the ``mass" profile is motivated by recent arguments that high-$z$ massive star-forming disk galaxies exhibit, on average, an outer falloff velocity profile (\citealt{vanDokkum2015, Genzel2017, Lang2017}).
We note that the ``mass'' profile in GalPak$^{\rm 3D}$ is mathematically different from the general formulation of the velocity profile of a Freeman disk (e.g., \citealt{Freeman1970, Binney2008}), which has been used in several studies of high-$z$ galaxies  (e.g.  \citealt{ForsterSchreiber2006, Gnerucci2011a, Ubler2018}).
However, the ``mass'' profile provides a good approximation, so that we tested the falloff with this profile.

In GalPAK$^{\rm 3D}$,  a total of 10 free parameters are used for fitting in the case of a tanh profile, including the $x_c$, $y_c$, $z_c$ positions, the disk half-light radius ($r_e$), the total flux $f_{\rm tot}$, the inclination $i$, position angle PA, the turnover radius $r_{\rm t}$, the maximum rotational velocity $V_{\rm max}$, and the one-dimensional intrinsic dispersion $\sigma_{0}$.
On the other hand, nine parameters are constrained for the mass profile, excluding the turnover radius.
We assumed an exponential disk (with S{\'e}rsic index $n=1$) for the flux distribution, which can be applied for main-sequence galaxies (e.g. \citealt{Wuyts2011a}).
With varying parameters, a simulated galaxy model is convolved (in 3D) with the PSF (or the synthesized beam) and the instrumental line-spread function (LSF) specified for each instrument (which corresponds to spectral resolution in our case) to fit the observed galaxy. 
The remaining setup for the model is arranged to default; the velocity dispersion is calculated as a ``thick" disk, and the vertical flux distribution ($I(z)$) is assumed as  a Gaussian in defining the characteristic thickness of the disk.
The $random\underbar{ }scale$ that sets the width of the proposal distribution for the next new set of parameters is chosen to be between 7 and 15 to get an MCMC acceptance rate between 20\% and 50\%, which was the recommended value in the instructions for GalPaK$^{\rm 3D}$.
The number of samplings (i.e., iteration) used for the model construction is 25,000 at first and increased to as high as 55,000 if necessary.
In each test run, we changed the boundaries for the maximum circular velocity to be high as $1500$~km s$^{-1}$ to obtain reasonable constraints of the model.
We used the data cube with the spectral resolution of 40~km s$^{-1}$ to ensure the spectral resolution and the S/N per channel to be sufficient to model the galaxy.

\begin{figure}[tb]
\centering
\includegraphics[width=8cm, bb=0 0 1200 1000]{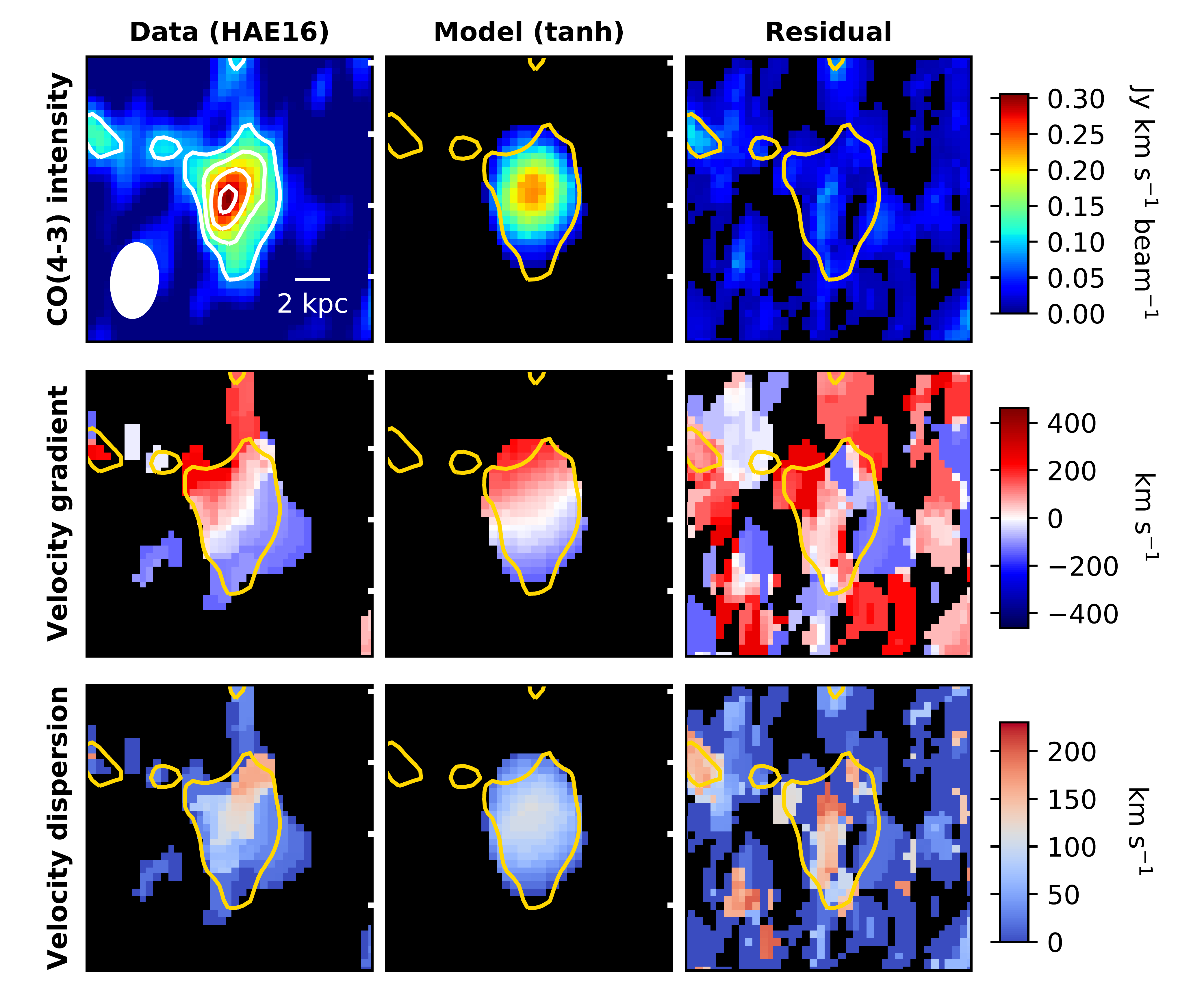}
\caption{Intensity, velocity gradient, and dispersion maps of the observed (first column), modeled (second column, convolved with the beam), and the residual (third column) data cubes of HAE~16 based on the tanh profile. Each row, from top to bottom, shows the CO~(4--3) intensity, velocity gradient, and velocity dispersion maps, respectively. 
For data, the velocity gradient and velocity dispersion maps are clipped below the flux below $1.8~\sigma$ for better representation of the maps when making the moment maps. For the model and residual, $1.3~\sigma$ is clipped for all moment maps. In the top-left panel, the contours are drawn from the intensity map starting from $3~\sigma$ in steps $2~\sigma$, i.e., $3~\sigma, 5~\sigma, 7~\sigma$, ... Only the $3~\sigma$ contour is shown in the remaining panels. Each panel has a width of $2''$. The physical scale of 2 kpc is shown in the top-left panel. The beam size of the CO~(4--3) data ($0''.52\times0''.32$) is shown by white filled ellipse. \label{fig:hae16_tanh}}
\end{figure}

\begin{figure}[tb]
\centering
\includegraphics[width=8cm, bb=0 0 1200 1000]{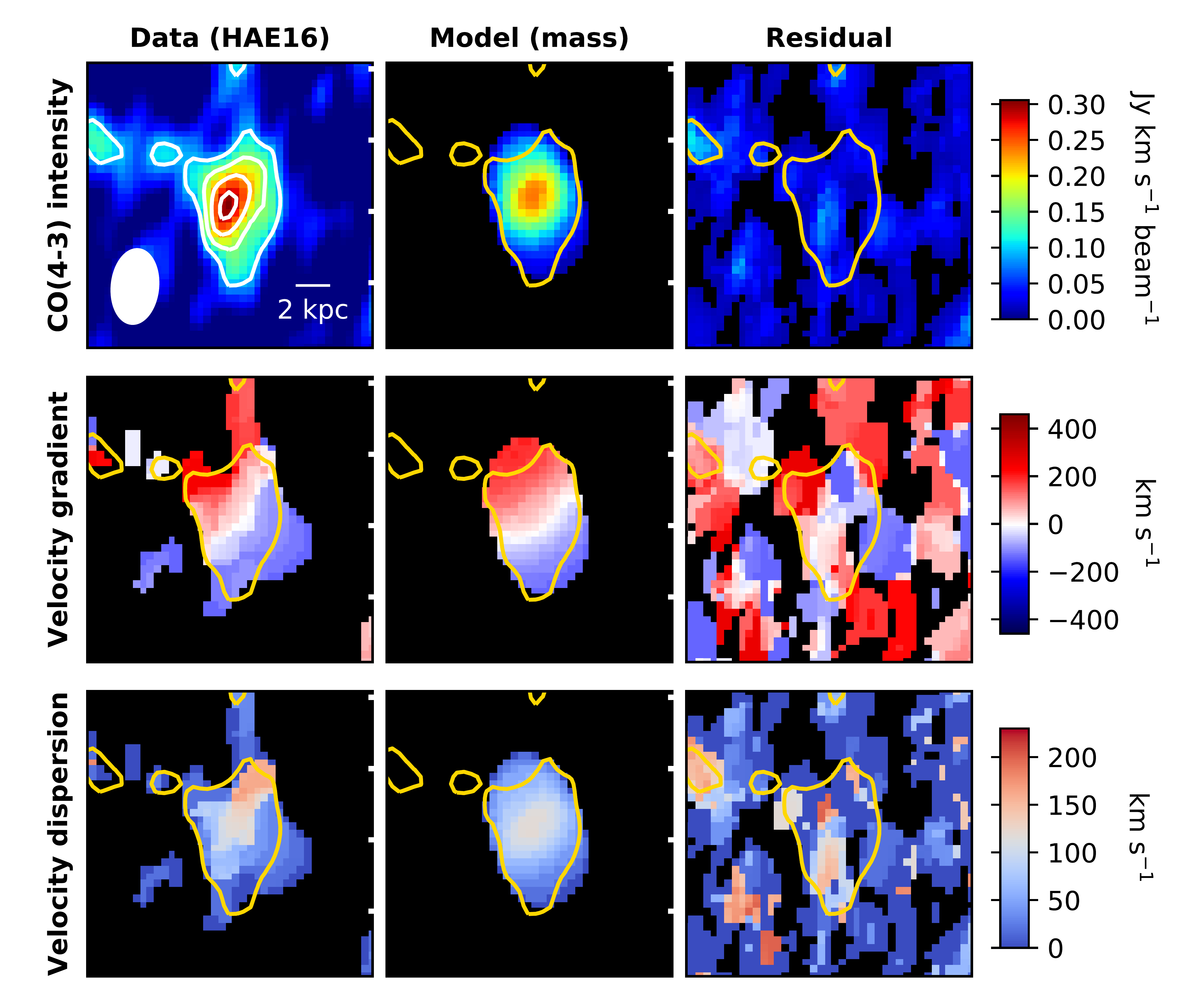}
\caption{Intensity, velocity gradient, and dispersion maps of the observed (first column), modeled (second column) and the residual (third column) data cubes of HAE~16 based on the mass profile. See the description in Fig~\ref{fig:hae16_tanh}.\label{fig:hae16_mass}}
\end{figure}

For HAE~16, the MCMC chains are converged with the reduced $\chi^{2}$ value of $\chi^{2}_{\rm reduced} \approx 1.37$ for both profiles.
The $\chi^{2}_{\rm reduced}$ values are close to unity, suggesting that both velocity profiles are capable of explaining the observed data sets for HAE~16 at a given spatial resolution.
The best-fit models for HAE~16 with different assumptions of velocity profiles are presented in Fig~\ref{fig:hae16_tanh} and \ref{fig:hae16_mass} with thes best-fit parameters in Table~\ref{tab:modelparams1}.

\begin{table}[tb]
\caption{Best-fit parameters from GalPaK$^{\rm 3D}$\label{tab:modelparams1}}
\begin{center}
\begin{tabular}{ccc}
        &               &       HAE~16  [95\% CI] \\  \hline \hline 
        &$r_e$ (kpc)    &       $2.85$ [$2.74,2.99$] \\ 
        &$V_{\rm max}$ (km s$^{-1}$) & $433$ [386,494] \\ 
tanh    &$\sigma_0$ (km s$^{-1}$)&    $76$ [69,83]\\ 
        &$i$ (degree)    &       $38$ [33,43] \\ 
        &$PA$ (degree)    &       $17$ [14,19] \\ 
        &$r_t$ (kpc)   &       $1.98$ [$1.77,2.19$] \\ 
\cline{2-3}
&$v/\sigma_0$ &  $5.7 \pm 1.0$ \\ \hline
        &$r_e$ (kpc)    &       $2.80$ [$2.64,2.94$] \\ 
        &$V_{\rm max}$ (km s$^{-1}$) & $301$ [290,317]\\ 
mass    &$\sigma_0$ (km s$^{-1}$)&    $84$ [76,90]\\ 
        &$i$ (degree)    &       $47$ [44,50] \\ 
        &$PA$ (degree)    &       $22$ [20,25]\\ 
\cline{2-3}
&$v/\sigma_0$ &  $3.6 \pm 0.4$\\ \hline
\end{tabular}
\end{center}
\tablecomments{$r_e$: the disk half-light radius, $V_{\rm max}$: the maximum rotational velocity, $\sigma_{0}$: the one-dimensional intrinsic dispersion, $i$: the inclination, $PA$: position angle, $r_{\rm t}$: the turnover radius}
\end{table}%

\subsection{Best-fit parameters}\label{sec:diskresults}
The derived range of maximum rotational velocity, $V_{\rm max} = 301$ and $433$~km s$^{-1}$ (for the mass and tanh profiles, respectively), suggests that HAE~16 has a slightly higher rotational velocity than what has been observed in field galaxies with a mean value of $\sim 200$ km s$^{-1}$ at $z\sim2$ for those with comparable mass in H$\alpha$ emissions(e.g.  \citealt{ForsterSchreiber2009,FS2018a, Wisnioski2015}) and in CO emissions (e.g., \citealt[see also Figure~\ref{fig:fwhm}]{Tacconi2013, Daddi2015, Aravena2019, Bourne2019}.
We note that the FWHMs of the CO lines from 1D Gaussian fitting are also high, i.e., $452\pm101$ km s$^{-1}$ and $547\pm138$~km s$^{-1}$ for the CO~(4--3) tapered map and the CO~(3--2) map, respectively, regardless of the assumption of velocity profile.

There are a few limitations suggested by the inspection of the best-fit parameters.
While the maximum rotational velocity may be larger than that of field galaxies, the maximum rotational velocity largely depends on the assumption of different velocity profiles.
The lack of sensitivity in the outer disk region is one of the main causes that introduce such a difference.
In addition, there is a covariance between the maximum rotation velocity and inclination, which shows the tightest negative correlation among the joint distributions of the parameters.
The maximum rotational velocity $V_{\rm max}$ is higher for the tahn profile, where the inclination is lower than that of the mass profile.
We need both deeper and higher angular resolution imaging observations for HAE~16 to constrain the velocity profile and the maximum velocity without any degeneracies.
We also note the estimated size of the galaxy is consistent with each other, which is $2.8\pm0.1$ kpc, which is slightly larger than the image-based deconvolved FWHM ($1.8\pm0.7$ kpc). 
We note that the FWHM from the gaussian fitting corresponds to $n=0.5$ case, which is different from what we assumed for the kinematical modeling ($n=1$).
The errors are not measured in the same way, and the systematic errors between different fitting may contribute to the different estimates. 
But, all of these values are still marginally smaller than the typical CO sizes for field galaxies that we discussed in Section~\ref{sec:comparison}.
Bearing these degeneracies and uncertainties in mind, we discuss the nature of HAE~16 with the best-fit model parameters and taking the difference as systematic errors.

\subsection{A rotation-dominated galaxy : high $V/\sigma$}
\begin{figure}
\centering
\includegraphics[width=0.48\textwidth, bb=0 0 1000 1000]{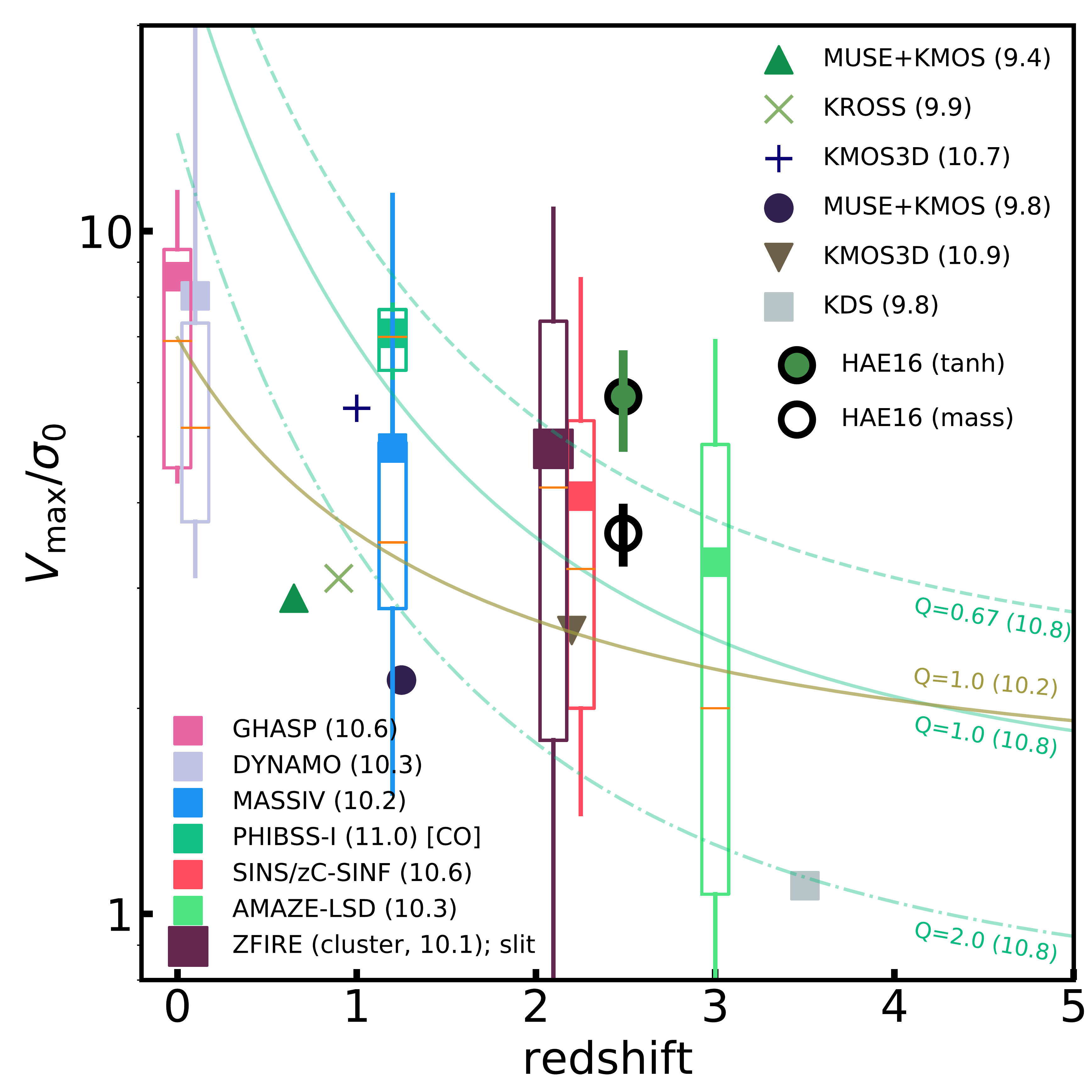}
\caption{The value of $V_{\rm max}/\sigma_0$ using the best-fit results of the HAE~16 kinematic modeling from GalPaK$^{\rm 3D}$ obtained from different velocity profiles using the ``tanh" and ``mass" profiles. $V_{\rm max}/\sigma_0$ against redshift for the comparison samples spanning $0 < z < 4$ in fields and clusters. 
The number next to each survey name in the legend corresponds to the mean stellar mass of the galaxies probed in log scale; $V_{\rm max}/\sigma_0$ also depends on stellar mass. 
All surveys other than ZFIRE are obtained by integral-field unit (IFU) observing [O$\textsc{III}$], [O$\textsc{II}$] or H$\alpha$, or by millimeter interferometry observing CO emission lines. 
ZFIRE is obtained by (multiobject) slit spectroscopy. 
We show the median values for MUSE+KMOS, KROSS, KMOS$^{\rm 3D}$ and KDS, which were compiled in \citet{Turner2017a}. 
GHASP, DYNAMO, MASSIV, PHIBSS-I, SINS/zC-SINF, AMAZE-LSD, and ZFIRE results are shown as boxes representing the central 50\% and the vertical lines for the 90\%. 
Filled squares associated with these boxes with the same colors show the mean values of $V_{\rm max}/\sigma_0$, and the horizontal orange lines are for the median. 
The solid and dashed lines are based on the expected evolution of $V_{\rm max}/\sigma_0$, assuming gas fraction and Toomre $Q$ parameter based on equation~\ref{eq:toomreQ}. 
For the gas fraction, we use the gas fraction scaling relation obtained in \cite{Genzel2015}, which is a function of the stellar mass, redshift, and deviation from the main sequence. 
Here, we assumed galaxies are on the main-sequence. 
Each curve shows the different Toomre $Q$ parameters with different stellar mass written within the parentheses.\label{fig:vsig}}
\end{figure}

How different is HAE~16 compared to other galaxies in fields and clusters in kinematical properties?
First, we evaluate how this disk galaxy is dominated by the ordered rotation.
The ratio between the rotation velocity and the intrinsic dispersion, $V_{\rm max}/\sigma_0$, provides a measure of the strength of the ordered rotation relative to random motions.
In general, a galaxy is rotation dominated when the ratio is larger than unity.
Within the uncertainties of the model constraints, the high values of $V_{\rm max}/\sigma_0 = 5.7$ (tanh) and $3.6$ (mass) suggest that the galaxy is rotation dominated, and the values are plotted in Figure~\ref{fig:vsig}.

For comparison, the results of field surveys are plotted in Figure~\ref{fig:vsig}, which used either H$\alpha$, [OIII] or [OII] lines using the IFU, or CO lines using millimeter interferometry.
For this plot, 
we used the results from MUSE+KMOS (\citealt{Swinbank2017}), KROSS (\citealt{Harrison2017}), KMOS$^{\rm 3D}$ (\citealt{Wisnioski2015}) and KDS (\citealt{Turner2017a}), which were compiled in the KDS paper. 
The remaining surveys of GHASP (\citealt{Epinat2008a, Epinat2008b}), DYNAMO (\citealt{Green2014}), MASSIV (\citealt{Epinat2012}), PHIBSS-I (\citealt{Tacconi2013}), SINS/zC-SINF (\citealt{ForsterSchreiber2009, FS2018a}), and AMAZE-LSD (\citealt{Gnerucci2011b}) are retrieved from the literature.

We compare with the SINS/zC-SINF survey, which probed field galaxies at a similar redshift and mass range.
It revealed that the average ratio of $V_{\rm max}/\sigma_0$ is 4.1 with the median of 3.2 (\citealt{FS2018a}).
Based on these values, we can conclude that HAE~16 shows a degree of rotation similar to or slightly higher than that of average field population, though the measurements are based on H$\alpha$, instead of CO. 
Though we need to verify whether the ionized gas and the molecular gas trace the same gas dynamics, they may not be significantly different globally for field galaxies at least (\citealt{Ubler2018}).

For another comparison, we take the values of $V_{\rm max}/\sigma_0$ for cluster galaxies from the results of ZFIRE (\citealt{Alcorn2018}), which is a slit spectroscopic survey using Keck with a subset of cluster galaxies at $z\sim2$.
We note that the ZFIRE cluster galaxies are even less massive than comparable field populations and HAE~16.
This is the only available literature for (proto)cluster galaxies at $z\gtrsim2$ that allow for a comparison, providing the largest sample available.
Interestingly, the average $V_{\rm rot}/\sigma_0$ for these less massive galaxies in a $z\sim2$ cluster is also slightly higher than that for $z\sim2$ field galaxies, but with a large scatter.

We note that the ratio tends to be lower for less massive galaxies and for those at higher redshifts, which is known as kinematical downsizing (e.g.  \citealt{Simons2016}).
Based on the high value (especially for the tanh profile) found in HAE~16 and the ZFIRE clusters, it is a tantalizing trend that suggests earlier and perhaps rapid formation of ``disk" galaxies in high-$z$ clusters, such that kinematical downsizing in the overdense regions of high-$z$ (proto)clusters precedes that in general fields, along with hierarchical structural formation as expected.
However, the uncertainty in our estimate (e.g., the lower value from the mass profile) makes this less convincing.
We need a larger number of samples and deeper observations to confirm this.

Given the constraints of $V_{\rm max}$, $\sigma_0$ and the gas fraction $f_{\rm gas}$, we can calculate the
Toomre $Q$ parameter ($Q_{\rm Toomre}$; \citealt{Toomre1964}) to constrain the stability of the (rotation-dominated) disk.
The Toomre-$Q$ parameter is expressed as
\begin{eqnarray}\label{eq:toomreQ}\nonumber
Q_{\rm gas} =& \frac{\sigma_0 \kappa}{\pi G \Sigma_{\rm gas}} = \Big(\frac{\sigma_0}{V_{\rm max}}\Big) \Big(\frac{a(V_{\rm max}^2 R_{\rm disk}/G)}{\pi R_{\rm disk}^2 \Sigma_{\rm gas}}\Big) \\ 
= &\Big(\frac{\sigma_0}{V_{\rm max}}\Big) \Big(\frac{a M_{\rm tot}}{M_{\rm gas}}\Big) = \Big(\frac{\sigma_0}{V_{\rm max}}\Big) \Big(\frac{a}{f_{\rm gas}}\Big),
\end{eqnarray}
 where $\kappa$ is the epicyclic frequency, $\Sigma_{\rm gas}$ is the gas surface density, $G$ is the gravitational constant, $R_{\rm disk}$ is the disk radius, and $M_{\rm tot}$ is the total mass. 
We use the baryonic mass ($M_{\rm tot} = M_{\rm gas} + M_{\rm star}$) for now.
We use $a=\sqrt{2}$ for a disk with constant rotational velocity.
A simple calculation based on the best-fit $V_{\rm max}/\sigma_0$ and gas fraction gives $Q_{\rm gas}\ll 1$, ranging between $0.3$ and $0.5$, and thus an even lower value than the critical $Q_{\rm crit}$ applied for a thick disk (\citealt{Goldreich1965b}).

This implies that the galaxy is gravitationally unstable, even though the galaxy shows a disk-like feature with the a smooth velocity gradient and high $V_{\rm max}/\sigma_0$ value. 
The gravitationally unstable disk may be a natural consequence of the gas-rich disk.

In Figure~\ref{fig:vsig}, we show a toy model in which the redshift evolution of $V_{\rm max}/\sigma_0$ is explained by the gas fraction evolution assuming a marginally stable disk, i.e., Toomre $Q$ to be close to 1 (\citealt{Wisnioski2015}).
HAE~16 slightly deviates from this model, such that the linear Toomre-stability analysis may no be longer valid. 
This recalls a disk with violent gravitational instability (e.g., \citealt{vandenBergh1996, Elmegreen2005, Bournaud2008, Genzel2008, Agertz2009, Ceverino2010, Genel2012}). 
In this scenario, the disk can no longer support the rotation by itself and fragments into clumps.
This is consistent with the Kp-band image (\citetalias{minju2017a}), which shows clumpy structures.
Such clumps may lead the gas (clumps) to fall into the central region, creating a dense stellar component in the near future to stabilize the structure.
The smaller size of the CO~(4--3)-emitting region may also support this picture and a large amount of gas might already have transferred into the central region.

\bibliographystyle{aasjournal}
\bibliography{minjujournal}



\end{document}